\documentclass[twocolumn]{emulateapj}
\usepackage{graphicx}
\newcommand{\be}{\begin{equation}}
\newcommand{\ee}{\end{equation} }
\newcommand{\ba}{\begin{eqnarray}}
\newcommand{\ea}{\end{eqnarray}}

\newcommand{\bmu}{\mbox{\boldmath$\mu$}}
\newcommand{\bOmega}{\mbox{\boldmath$\Omega$}}
\newcommand{\nn}{\mbox{} \nonumber \\ \mbox{} }
\newcommand{\kB}{k_{\rm B}}
\newcommand{\K}{^\circ{\rm K}}

\shorttitle{Giant Primeval Magnetic Dipoles}
\shortauthors{Thompson}	
\begin{document}
\title{Giant Primeval Magnetic Dipoles}
\author{Christopher Thompson}
\affil{Canadian Institute for Theoretical Astrophysics, 60 St. George St., Toronto, ON M5S 3H8, Canada.}		

\begin{abstract}
Macroscopic magnetic dipoles are considered as cosmic dark matter.  Permanent
magnetism in relativistic field structures can involve some form of superconductivity, one example
being current-carrying string loops (`springs') with vanishing net tension.
We derive the cross section for 
free classical dipoles to collide, finding it depends weakly on orientation when mutual precession 
is rapid.   The collision rate of `spring' loops with tension ${\cal T} \sim 10^{-8}c^4/G$
in galactic halos approaches the measured rate of fast radio bursts (FRBs) if the loops comprise most 
of the dark matter.  A large superconducting dipole (LSD) with mass  $\sim 10^{20}$ g
and size $\sim 1$ mm will form a $\sim 100$ km
magnetosphere moving through interstellar plasma.  Although hydromagnetic drag is generally weak,
it is strong enough to capture some LSDs into long-lived rings orbiting supermassive black holes 
(SMBHs) that form by the direct collapse of massive gas clouds.  
Repeated collisions near young SMBHs could dominate the global collision rate, thereby broadening 
the dipole mass spectrum.   Colliding LSDs produce tiny, hot electromagnetic explosions.  The accompanying 
paper shows that these explosions couple effectively to propagating low-frequency electromagnetic modes, 
with output peaking at 0.01-1 THz.  We describe several constraints on,
and predictions of, LSDs as cosmic dark matter.  The shock formed by an infalling LSD
triggers self-sustained thermonuclear burning in a C/O (ONeMg) white dwarf (WD) of mass $\gtrsim 1\,M_\odot$
($1.3\,M_\odot$).  The spark is generally located well off the center of the WD.
The rate of LSD-induced explosions matches the observed rate of Type Ia supernovae.
\end{abstract}

\section{Introduction}\label{s:intro}

We consider macroscopic magnets as a component of the cosmic dark matter, showing that the hypothesis has a
number of interesting observational consequences.   The underlying theoretical motivation, although slender, goes
back to the demonstration that cosmic strings may carry substantial electric currents \citep{witten85}.
In some circumstances the magnetic field winding around the string grows strong enough to compensate the
relativistic string tension, in which case macroscopic static loops of the string would survive indefinitely as a form of 
kinematically cold dark matter \citep{otw86,copeland87,haws88,davis89}.
But the possibility of relativistic field structures carrying a magnetic dipole moment is a more general one.

Our interest here is in the interactions between these large superconducting dipoles (LSDs) which, although 
individually rare, can collectively be quite frequent.   
Direct collisions are a source of intense, but
narrow, electromagnetic pulses, details of which are examined in the companion paper (\citealt{t17}, hereafter Paper I).
In fact, the motivation for this investigation was provided by the 
discovery of rapid and intense radio bursts (FRBs;  \citealt{lorimer07,thornton13}) occurring at a rate of 
about $3\times 10^3$ per day above a fluence threshold of $\sim 5\times 10^{-17}$ erg cm$^{-2}$ \citep{crawford16},
or $1-3\times 10^4$ per day as inferred from the one demonstrated
repeating source \citep{scholz16}.  The pulses are highly dispersed, suggesting a cosmological origin,
which has recently been confirmed indirectly by a measurement of Faraday rotation \citep{ravi16}, and 
directly by localization \citep{chatterjee17,marcote17,tendulkar17}.
Then the energy emitted in the GHz band reaches at least $10^{39}$ erg in some cases (unless the
emission is highly beamed: \citealt{lg14,katz16}).

We show here that two channels are available for LSD collisions.  Very rarely LSDs orbiting in 
galactic halos will collide with each other.   A small subset of LSDs can also be captured gravitationally
by supermassive black holes (SMBHs) if the holes form by direct collapse of dense gas clouds.   The collision rate
through the first channel approaches the (still roughly) measured FRB rate if the LSDs
i)  comprise a significant fraction of the cosmic dark matter;  ii) have a large enough mass
to power a FRB, after allowing for a $\sim 1$-$10\%$ efficiency of GHz emission (Paper I); and iii)
have magnetic fields not far below the electroweak scale ($10^{24}$ G).  In the `spring' model conditions ii) and iii)
implyd a string tension ${\cal T} \sim 10^{-8} c^4/G$.

Each LSD orbiting through a galaxy forms a small magnetosphere and experiences very weak drag off the
surrounding plasma.  However, the drag effect is more significant for the LSDs that are adiabatically
compressed close to a newly formed SMBH following its collapse.   These dipoles settle gradually toward 
the innermost stable circular orbit (ISCO), just outside of which the drag force vanishes.   We show that collisions
significantly broaden their mass distribution, allowing for a range of radio pulse energies starting
from a peaked primordial mass distribution.  In this dense environment, a large fraction of the dipoles are removed by
collisions.   The companion paper also shows that a GHz pulse can escape this environment if the black hole has entered
a phase of weak and radiatively inefficient accretion.

A significant contribution of LSDs to the dark matter just above the mass range considered here ($\sim 10^{20}$ g) 
is strongly constrained by sub-pixel microlensing \citep{niikura17}.  Destructive collisions with white dwarfs (WDs) have
also been considered as a limitation on the abundance of primordial black holes around this mass 
\citep{capela13,graham15}.  A LSD that is gravitationally deflected into a WD will produce a similarly 
localized heating.  We calculate the critical WD mass (central density) above which this heating triggers
self-sustained thermonuclear burning, which turns out to depend on the WD composition, and consider whether 
the heated zone is large enough to drive a direct detonation.  The calculated rate of thermonuclear explosions
is close to the observed Type Ia rate, and there is an interesting
prediction of a depletion of old and massive WDs in the Solar neighborhood.   

The existence of the LSDs considered here does not strictly imply the presence of a network of horizon-crossing cosmic
strings or oscillating string loops.   Nonetheless existing bounds on local string networks from pulsar timing
(e.g. \citealt{lentati15}) are consistent with the range of tensions considered here.   Strong gravitational
lensing by the long string \citep{ks84} would be restricted to sub-arcsecond angular scales.   

The web of physical processes described here and in Paper I is applied to the repeating FRB 121102.  Gravitational
lensing by a $\sim 10^6\,M_\odot$ SMBH naturally generates a minimum time delay of $\sim 30$ s.  We show that reflection of
a strong electromagnetic pulse by dense plasma clouds is restricted to larger distances from the black hole, corresponding
to time delays of hundreds of seconds.   Both of these timescales are present in the observed activity of FRB 121102
\citep{spitler16,scholz16}.  The pulses emerging from close to a SMBH are expected to show strong Faraday rotation,
or even be depolarized at finite bandwidth.

A key prediction of the model is 0.01-1 THz pulses even brighter and narrower than FRBs (Paper I).  High-frequency
measurements directly probe the underyling field structure if FRBs are powered by `tiny' electromagnetic explosions.
High-frequency pulses can escape from relatively dense plasma clouds around SMBHs, and are a possible source of `noise'
in CMB experiments.  Detecting them would discriminate between emission from plasma structures
of a size larger than a radio wavelength $\lambda$, versus relativistic field structures smaller than $\lambda$.  
One expects the emission to be weaker at longer wavelengths if the emitting plasma is comparable in size to a neutron
star.

The redshift distribution
of electromagnetic pulses from colliding LSDs depends on whether the collision rate is dominated by compact rings
orbiting SMBHs, or instead by rare collision events distributed throughout galactic halos.  In the first case, the FRB
emission is weighted toward higher redshifts, closer to the epoch of collapse of the first SMBHs.  In addition, only SMBHs
which form directly from gas collapse and do not merge with other SMBHs are potential emission sites of FRBs.  We show
that the LSD ring around a SMBH does survive repeat injections of very dense plasma from the tidal disruption of stars.

Finally, the nearest LSD is predicted to lie about 20 AU from the Sun, and move with extreme proper motion with
a residency time of $\sim 0.3$ yr.  Scaling down from planetary magnetospheres suggest that the $\sim 100$ km zone
around the nearest LSD is a possible $\mu$Jy source of coherent radio emission.

The plan of this paper is as follows.  The cross section for the collision of free classical magnetic dipoles is presented
in Section \ref{s:dipole}, with further details in the Appendix.   The interaction of a moving LSD with ambient plasma 
is described in Section \ref{s:plasma}, including trapping near SMBHs.   The rate of collisions between LSDs bound
in galactic halos is calculated in Section \ref{s:rate}.   The identification of LSDs with GUT-scale superconducting
`spring' loops is considered in Section \ref{s:scs}.  The collisional evolution of the dipole number, mass spectrum,
and radiation energy spectrum of a LSD ring near the ISCO of a SMBH is calculated in Section \ref{s:smbh}.
Details of the interaction with realistic model WDs can be found in Section \ref{s:wd}.
The concluding Section \ref{s:summary} describes various tests, observational consequences, and predictions of
LSD dark matter and LSD collisions as the mechanism of FRBs.

\section{Collision Cross Section of Classical Magnetic Dipoles (in Vacuum)}\label{s:dipole}

Here we calculate the cross section for two classical magnetic dipoles to collide.  
The dipoles are otherwise unbound and assumed to propagate in a vacuum.  When the magnetic
field contributes a significant part of the inertia of each dipole, ambient plasma
screens the magnetic field only on much larger scales than the collision impact parameter.

The relation between magnetic dipole moment $\mu$ and rest mass ${\cal M}$ is normalized to
\be\label{eq:dipmass}
f_{\rm em} {\cal M}c^2 = \mu^2 {\cal R}^{-3},
\ee
where ${\cal R}$ is a characteristic radius.  Setting the coefficient $f_{\rm em}$ on the left-hand side to unity 
gives the electromagnetic inertia of a uniformly magnetized sphere of radius ${\cal R}$, 
in which case the relation between mass and radius is\footnote{Throughout 
this paper we use the shorthand $X = X_n \times 10^n$, where quantity $X$ is expressed in c.g.s. units.}
${\cal M} \sim 3\times 10^{19}\,B_{22}^2 {\cal R}_{-1}^3$ g.   Additional (e.g. toroidal or scalar) field components 
are needed to stabilize the dipole, and so we expect $f_{\rm em} \lesssim 1$.   A concrete example is given 
in Section \ref{s:scs}.

Two colliding dipoles will merge if the $-\bmu\cdot{\bf B}$ energy grows in magnitude above
their initial kinetic energy.
Because the dipole-dipole interaction scales as $r^{-3}$, there is no centrifugal barrier to a direct
collision.  Then a first estimate of the collision impact parameter of two identical dipoles is 
\ba\label{eq:rcol}
b_{\rm col} \sim R_{\rm col} &\equiv& \left({\mu^2\over {\cal M}v^2}\right)^{1/3} \nn
 &=& f_{\rm em}^{1/3} \left({v\over c}\right)^{-2/3} {\cal R} = 10\,f_{\rm em}^{1/3}\,v_{7.5}^{-2/3}{\cal R}_{-1}
\quad {\rm cm}.\nn
\ea
Here $v$ is the velocity at infinity of each dipole.  This cross section depends on
the dipole size, but not individually on $\mu$ and ${\cal M}$.

The initial orientation of the dipoles does not matter as long as each dipole precesses rapidly in
the magnetic field of the other.   The precession frequency of one dipole in
the magnetic field of a second is
\be
\omega_g = \left( |{\bmu}_1\times {\bf B}_2({\bf r}_{12})|\over {\cal I}_1 \right)^{1/2},
\ee
where ${\bf r}_{12} \equiv {\bf r}_1 - {\bf r}_2$ is the separation.  This works out to
\be
{\omega_g(R_{\rm col})\over v/R_{\rm col}} \sim f(\mu,v) \equiv
\left({{\cal M} R_{\rm col}^2\over {\cal I}}\right)^{1/2} \sim {R_{\rm col}\over {\cal R}}
\ee
at the capture impact parameter $r_{12} \sim R_{\rm col}$.  The fastness parameter $f$ can be written as
\ba\label{eq:fast}
f(\mu,v) &=& f_{\rm em}^{1/3}\left({v\over c}\right)^{-2/3}\left({{\cal I}\over {\cal M}{\cal R}^2}\right)^{-1/2}\nn
 &=& 100\,f_{\rm em}^{1/3}v_{7.5}^{-2/3}\left({{\cal I}\over {\cal M}{\cal R}^2}\right)^{-1/2}.
\ea
Here ${\cal I}$ is the moment of inertia of a dipole, whose rotational degrees of freedom are reduced to those
of a sphere.

\begin{figure}
\epsscale{1.1}
\plotone{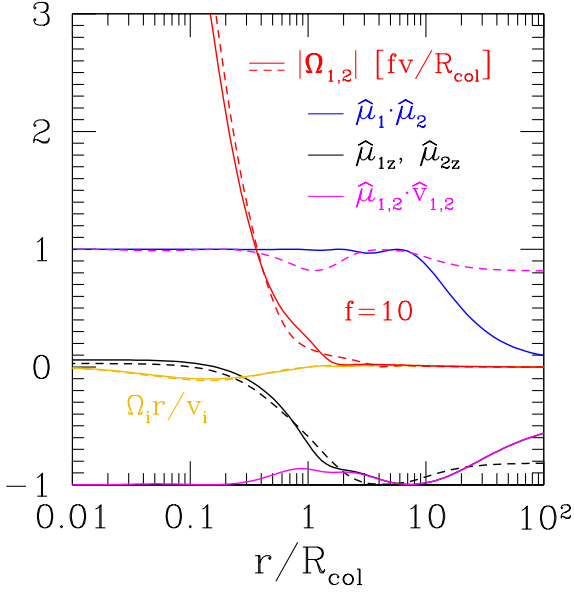}
\vskip .1in
\caption{Two classical magnetic dipoles collide along the $z$-axis with vanishing initial rotation,
and fastness parameter (\ref{eq:fast}) $f = 10$.
Black lines:  projection of unit magnetic moment onto the collision axis.  Blue line:
relative orientation of the two moments.  Magenta lines:  projection ${\hat\mu}_i\cdot{\hat v}_i$
of magnetic moment $i = 1,2$ on velocity, which is nearly radial inside the collision radius
$R_{\rm col}$.  Red lines:  dipole spin vectors in units of $f v/R_{\rm col}$, where $v$ is
velocity of each dipole at infinity.  Gold lines:  spin frequency multiplied by the collision time.}
\vskip .2in
\label{fig:collision}
\end{figure}

In the regime of rapid precession ($f \gg 1$), two dipoles reaching a separation $\lesssim R_{\rm col}$ quickly rotate into 
a configuration with each magnetic moment nearly aligned parallel (anti-parallel) to its radial velocity,
${\hat v}\cdot{\hat \mu} = \pm 1$, and ${\hat\mu}_1\cdot{\hat\mu}_2 = 1$ (Figure \ref{fig:collision}).  
This generates an attractive force and minimizes the torque, the precession rate, and the interaction energy 
\be
{\cal E}_{12} = {\bmu_1\cdot\bmu_2 - 3(\bmu_1\cdot\hat r_{12})(\bmu_2\cdot\hat r_{12})\over r_{12}^3}
=-{2\bmu_1\cdot\bmu_2\over r_{12}^3}.
\ee

\begin{figure}
\epsscale{1.1}
\plotone{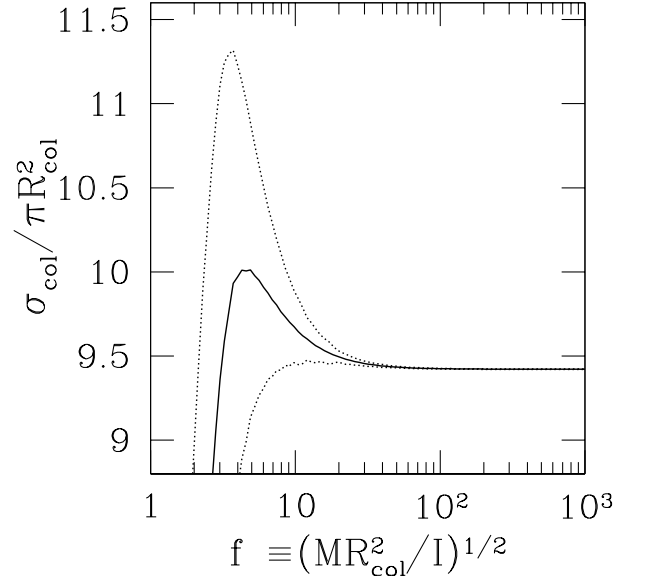}
\vskip .1in
\caption{Solid line:  mean cross section for the collision of two classical and non-relativistic 
dipoles in vacuum, as a function of the fastness parameter $f \equiv ({\cal M}R_{\rm col}^2/I)^{1/2}$.
Dotted curves show the variance in the result for randomly oriented magnetic moments, with
vanishing initial spins.}
\vskip .2in
\label{fig:cs}
\end{figure}

The generalization to dipoles with unequal magnetic moments and masses is straightforward.  The collision
radius is now defined as 
\be\label{eq:rcolb}
R_{\rm col} \equiv \left[{2\mu_1\mu_2\over {\cal M}_r (\Delta v)^2}\right]^{1/3}.
\ee
Here ${\cal M}_r = {\cal M}_1{\cal M}_2/({\cal M}_1 + {\cal M}_2)$ is the reduced mass, and
$\Delta v$ is the relative velocity at infinity. This expression matches (\ref{eq:rcol}) when
$\mu_1 = \mu_2$ and ${\cal M}_1 = {\cal M}_2$, taking into account $\Delta v \rightarrow 2v$.
Complications with differing moments of inertia can be neglected when $f\gg 1$.

To evaluate the dispersion in capture cross section at finite $f$, we calculate numerically a large set of
trajectories with random orientations of $\bmu_i$ ($i = 1,2$) with respect to the collision axis
and vanishing initial spins.  Figure \ref{fig:cs} shows the result.  The cross section at infinite $f$ is
\be\label{eq:bcol}
b_{\rm col}(f=\infty) = 1.732\,R_{\rm col};  \quad \sigma_{\rm col}(f = \infty) = 9.42\, R_{\rm col}^2.
\ee
The variance in $\sigma_{\rm col}$ decreases to a few percent at $f = 10$, and asymptotes to zero as 
$f \rightarrow \infty$.  Notice that the collision rate depends weakly on the velocity dispersion of the 
dipoles, e.g. on the depth of the Galactic potential, since $\sigma_{\rm col} v \propto v^{-1/3}$.
Further details of the calculation, including a demonstration that the result depends only on $R_{\rm col}$
as defined in Equation (\ref{eq:rcolb}) when $f\gg 1$, can be found in the Appendix.

\subsection{Gravitational Effects}

The gravitational interaction between two dipoles must be taken into account when they move sufficiently
slowly, and the collision impact parameter $b_{\rm col}$ becomes large enough that $G{\cal M}^2/b_{\rm col}
\gtrsim \mu^2/b_{\rm col}^3$.   The critical speed above which the magnetic interaction dominates is
\be\label{eq:vmax}
{v\over c} \simeq 2.3\,f_{\rm em}^{-1/4}\left({G{\cal M}\over {\cal R}c^2}\right)^{3/4}
           = 1.0\times 10^{-5} f_{\rm em}^{-1/4} {\cal M}_{20}^{3/4} {\cal R}_{-1}^{-3/4}
\ee
in the center of mass frame.

\vfil\eject
\subsection{Post-Collision Relaxation}

As the two dipoles approach each other, conservation of $\bOmega_i\cdot\bmu_i$ implies
that the rotation and magnetic moment vectors becomes nearly orthogonal.  Annihilation cannot be complete
if $\mu_1 \neq \mu_2$, and a remnant dipole must be left behind following the collision.  The rotation
of this remnant can be estimated by balancing the orbital angular momentum with the final spin angular momentum,
${\cal M} R_{\rm col} v \sim {\cal I}\Omega_f$, giving a final spin frequency
\be
{\Omega_f {\cal R}\over c} \sim f_{\rm em}^{1/3}\left({v\over c}\right)^{1/3}.
\ee
Assuming that the two colliding dipoles did not have nearly equal masses, this remnant
rotational energy amounts to a fraction $\sim f_{\rm em}^{2/3}(v/c)^{2/3}$ of energy
liberated in the collision.\footnote{When $f_{\rm em} \ll 1$ one might imagine that additional
components of the stress-energy of the dipoles could resist annihilation and prevent collisional
energy release of the order of ${\cal M}c^2$.  In fact, $f_{\rm em} \simeq 0.025$ for 
GUT-scale superconducting `spring' loops.  Then the magnitude of the prompt energy release depends on the relative
sign of the currents on the colliding loops.   It is $\sim f_{\rm em}{\cal M}c^2$ for loops with
currents of like sign, and $\sim {\cal M}c^2$ for opposing signs.  See the discussion in Section \ref{s:scs}.}
It is lost to magnetic dipole radiation over the timescale
\be
t_{\rm mdr} \sim {{\cal I}c^3\over \mu^2\Omega_f^2} \sim f_{\rm em}^{-5/3} \left({v\over c}\right)^{-2/3} {{\cal R}\over c}
= {10^2\over f_{\rm em}^{5/3} v_{7.5}^{2/3}} {{\cal R}\over c},
\ee
some 2-4 orders of magnitude longer than the width of the prompt electromagnetic pulse.

\section{Interaction of LSD with Ambient Plasma}\label{s:plasma}

A single LSD moving through plasma at speed $v$ excludes most of the ambient charged 
particles out to a radius
\be\label{eq:rmag}
R_{\rm mag} \sim \left({\mu^2\over 4\pi n_{\rm ex} m_p v^2}\right)^{1/6}
= {{\cal R}\over (v/c)^{1/3}} \left({\bar\rho_{\rm LSD}\over m_p n_{\rm ex}}\right)^{1/6}.
\ee
Here $n_{\rm ex}$ is the ambient number density of free electrons, and
we define a characteristic density within a dipole radius ${\cal R}$,
\be\label{eq:barrho}
\bar\rho_{\rm LSD} \equiv {f_{\rm em}{\cal M}\over 4\pi{\cal R}^3} 
= 8\times 10^{21}\;f_{\rm em}{\cal M}_{20}{\cal R}_{-1}^{-3}\quad
{\rm g~cm^{-3}}.
\ee
This `magnetospheric' radius is much larger than the collision impact parameter
(\ref{eq:bcol}), meaning that the vacuum approximation is well justified
when treating the interaction between dipoles.  In the case of a LSD moving through
the interstellar medium at a speed $v \sim 10^{-3}\,c$, one finds $R_{\rm mag}
\sim 400\,f_{\rm em}^{1/6}{\cal M}_{20}^{1/6}{\cal R}_{-1}^{1/2}$ km.

The hydromagnetic drag force acting on the LSD only slightly perturbs its orbit, 
unless $n_{\rm ex}$ is very high compared with a typical interstellar particle density.   Writing
\be\label{eq:fdrag}
F_{\rm drag} = {1\over 2} C_d n_{\rm ex} m_p v^2 \cdot \pi R_{\rm mag}^2
\ee
in terms of a drag coefficient $C_d = O(1)$ (e.g. \citealt{bh72}) one finds a drag time
\ba\label{eq:tdrag}
t_{\rm drag} &=& {{\cal M} v\over F_{\rm drag}} \sim {8\over C_d f_{\rm em}} 
\left({\bar\rho_{\rm LSD}\over\rho_{\rm ex}}\right)^{1/2}{R_{\rm mag}\over c}\nn
      &=& 2\times 10^{13}\,f_{\rm em}^{-1/3}{\cal M}_{20}^{2/3}{\cal R}_{-1}^{-1}
           n_{\rm ex,0}^{-2/3}v_{7.5}^{-1/3}\quad{\rm yr}.
\ea
The large majority of LSD orbits within galactic halos can therefore be treated as collisionless.
Hydromagnetic drag clearly has a stronger effect on the rare LSD that happens to be accreted by
a star (Section \ref{s:wd}).

\subsection{Trapping by Supermassive Black Holes}

The hydromagnetic drag force just described opens up a new channel for trapping LSD near 
SMBHs, while avoiding accretion through the horizon.

A LSD orbiting a black hole experiences hydrodynamic drag even if its orbital
angular momentum is aligned with the angular momentum of the gas accreting onto the hole.
That is because a negative pressure gradient within the gas partly compensates the central
gravitational field and causes the gas to rotate more slowly than a Keplerian orbit.
Then an LSD in a circular orbit around a black hole of mass $M_\bullet$, with semi-major axis 
$a$ and angular frequency $\Omega(a) = (GM_\bullet/a^3)^{1/2}$, experiences a headwind
\be\label{eq:headwind}
\Delta v = \Omega a - \left(\Omega^2 a^2 + {1\over \rho_g}{dP_g\over \ln a}\right)^{1/2}
                  \simeq -{1\over 2\rho_g \Omega}{dP_g\over da}.
\ee

This headwind dies away where the radial pressure gradient vanishes, which happens just
outside the ISCO as defined for material particles without drag.
As a result, LSDs which experience strong enough drag to reach the inner part of the accretion
flow will tend to collect in a narrow ring just outside the ISCO.

We now show that hydromagnetic drag acts on an inspiraling LSD over timescales intermediate between
the orbital period and the age of the black hole.  This means that the LSD is sensitive to
a long-term average of the accretion flow, and will survive rapid
($\sim$ orbital period) turbulent fluctuations in the gas density and velocity fields.   
The inspiral may be concentrated in episodes of higher-than-average accretion, 
whereas the interaction between dipoles near the ISCO is spread out over longer intervals
when the accretion rate is relatively low.

To simplify matters, we consider a radiatively inefficient and geometrically thick accretion flow 
with inflow speed $\alpha \cdot \Omega a$, where $\alpha$ is related to the usual viscosity coefficient
\citep{ss73}.  Then the gas density needed to support a mass inflow rate $\dot M(a)$ is
\be\label{eq:rhog}
\rho_g(a) = {\dot M(a)\over \alpha \cdot 4\pi \Omega a^3} = 
{\dot m(a)\over \varepsilon_{\rm rad} \alpha\kappa_{\rm es} a} \left({GM_\bullet\over a c^2}\right)^{1/2}.
\ee
Here $\kappa_{\rm es} \sim \sigma_T/m_p$ is the electron scattering opacity.  
The dimensionless accretion rate $\dot m = 1$ corresponds to the rate needed to power 
the Eddington luminosity $4\pi GM_\bullet c/\kappa_{\rm es}$ from an accretion flow with radiative
efficiency $\varepsilon_{\rm rad}c^2$ per unit mass.  
The radial dependence of $\dot m$ is introduced to allow for a combined
pattern of inflow and outflow, with plasma partly diverted from an accreting equatorial band into
an outflowing polar wind \citep{bb99,yuan12}.   The gravitational radius is 
$R_g \equiv GM_\bullet/c^2$, so that, e.g., $\Omega(a) = (c/R_g) (a/R_g)^{-3/2}$.

Substituting Equation (\ref{eq:rhog}) into Equation (\ref{eq:tdrag}) gives
\be\label{eq:tdrag2}
\Omega t_{\rm drag} \;\sim\; {6\times 10^8 \over (\Delta v/c)^{1/3}}
{f_{\rm hydro}{\cal M}_{20}^{2/3}\over 
\dot m(a)^{2/3}f_{\rm em}^{1/3} {\cal R}_{-1} M_{\bullet,7}^{1/3}}\left({a\over 6R_g}\right)^{-1/2},
\ee
or equivalently
\be\label{eq:tdrag3}
t_{\rm drag} \;\sim\; 2.5\times 10^3\, 
{f_{\rm hydro} ({\cal M}_{20}\,M_{\bullet,7})^{2/3}\over
\dot m(a)^{2/3}f_{\rm em}^{1/3}{\cal R}_{-1}}\left({a\over 6R_g}\right)^{7/6}\quad {\rm yr}.
\ee
Here $M_{\bullet,7}$ is the black hole mass in units of $10^7\,M_\odot$.
We factor out some properties of the accretion flow and the hydromagnetic interaction
by defining $f_{\rm hydro} \equiv (\varepsilon_{\rm rad,-1}\alpha_{-2})^{2/3} C_d^{-1}$.

Reconsidering now the effect of intermittent accretion on the trapped LSD ring, one recalls
that a moderate fraction of the mass of $\sim 10^6$-$10^7\,M_\odot$ SMBHs may be build up by the 
tidal disruption of stars from the surrounding nuclear cluster \cite{wang04}.   The plasma density
at the ISCO could easily increase by 6-7 orders during such an event, if the disrupted star had
not yet left the hydrogen-burning main sequence.  Nonetheless,
one sees from Equation (\ref{eq:tdrag2}) that the drag time would remain longer than the orbital period.
The essential feature of a pressure maximum near the ISCO would also be sustained during such
an event, preventing an orbital instability of the LSD ring.  

In the remainder of the paper, we first discuss the interaction between LSDs that are broadly 
distributed throughout galactic halos (Section \ref{s:rate}) before turning to address their 
collisions near SMBHs (Section \ref{s:smbh}).

\section{Collision Rate in Galactic Halos}\label{s:rate}

We now consider the self-interaction of a LSD component of the cosmic cold dark matter (CDM).
A first estimate shows that the optical depth to collisions {\it between} LSDs within a galaxy
is very small.  Consider a Milky Way type galaxy (halo mass $M_h \sim 1\times 10^{12}\,M_\odot$)
with dark matter mass density $\rho_d \sim 3\times 10^{-24}$ g cm$^{-3}$ at a galactocentric 
radius $R \sim 3$ kpc (e.g. \citealt{schaller16}).  Magnetic dipoles are assumed to contribute a uniform fraction
$f_{\rm LSD}$ of $\rho_d$.  Then the optical depth to collisions between two dipoles is
\be
\tau_{\rm col}(R) \sim \pi b_{\rm col}^2(\sigma) R {\rho_d(R)\over {\cal M}},
\ee
where $\sigma \sim 150$ km s$^{-1}$ is the dark matter velocity dispersion.  Substituting 
Equation (\ref{eq:bcol}) gives $\tau_{\rm col} \sim 1\times 10^{-18} f_{\rm LSD}f_{\rm em}^{2/3}
{\cal R}_{-1}^2{\cal M}_{20}^{-1}$.

\begin{figure}
\epsscale{1.1}
\plotone{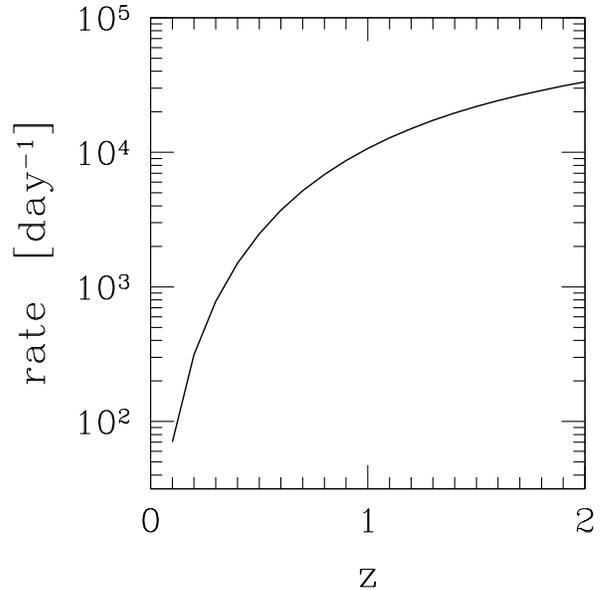}
\vskip .1in
\caption{Cumulative rate of collisions between LSDs comprising the entirety of the cosmic
dark matter ($f_{\rm LSD} = 1$), as a function of cosmological redshift.
LSDs are assumed to have uniform mass ${\cal M} = 10^{20}$ g, radius ${\cal R} = 0.1$ cm, 
and $f_{\rm em} = 0.025$ as appropriate to superconducting `spring' loops with tension
${\cal T} = 10^{-8}\,c^4/G$ (Section \ref{s:scs}).   The LSDs uniformly populate cold
dark matter halos with masses exceeding $10^{10}\,M_\odot$ in a CMB-normalized cosmology.  
Halo profiles are NFW fits calibrated by the simulations of \cite{klypin11} and \cite{prada12}.
No correction is made for additional concentration of dark matter toward the centers of
halos, resulting from the settling of gas.}
\vskip .2in
\label{fig:rate}
\end{figure}

Given that only a percent of the halo dark mass is concentrated this close to the center of the
galaxy, that the dark matter particles execute $\sim 10^3$ orbits over the Hubble time, and 
that 10 percent of the dark mass is bound in halos off mass $10^{12\pm 0.5}\,M_\odot$, one finds that a fraction 
$f_{\rm col} \sim \tau_{\rm col}(3~{\rm kpc})$ of the LSD bound in halos will suffer a 
binary collision within a Hubble time.  The rate out to a cosmological redshift $z\sim 1$ is
\ba\label{eq:colrate}
\Gamma_{{\rm col},z<1} &\sim& f_{\rm col} \cdot f_{\rm LSD}\Omega_{d,0}{c^3\over G{\cal M}}\nn
&\sim& 8\times 10^4 f_{\rm LSD}^2f_{\rm em}^{2/3}{\cal R}_{-1}^2 {\cal M}_{20}^{-2}\quad {\rm day}^{-1}\nn
\ea
in the standard $\Lambda$CDM cosmology with density $\Omega_{d,0} =  \rho_d(0)/\rho_{\rm cr}(0) = 0.25$ 
of dark matter relative to the critical density at $z = 0$ \citep{planck16}.

A more precise calculation of the collision rate is easily done by populating the universe
with model dark matter halos.  We use the mass distribution tabulated by \cite{klypin11} and the 
internal density profile of \cite{nfw97},
\be
\rho_d(r,M_h,z) = \rho_{\rm cr}(z)\,{\delta_c(M_h,z)\over (r/r_s)(1+r/r_s)^2},
\ee
as calibrated by \cite{prada12}.  Here $\rho_{\rm cr} = 3H(z)^2/8\pi G$, and the LSD space density is given by 
\be
n_{\rm LSD} = {f_{\rm LSD}\over {\cal M}} \rho_d.
\ee
The scale radius $r_s = r_{200}/c$, compactness $c$, and normalization $\delta_c = 200c^3/3[\ln(1+c) - c/(1+c)]$
are all functions of halo mass $M_h$ and redshift.  

Here we avoid the complications arising from gas flows within forming galaxies and their influence on the
central dark matter profile.   Radiative settling of gas has the effect of compressing
the dark matter near the centers of Milky-Way sized halos (e.g. \citealt{schaller16}), whereas persistent stirring
of the baryons may smooth out the dark matter cusp at the center of dwarf halos (e.g. \citealt{governato12}).  
The strength of this second effect has been disputed \citep{fattahi16}.   The overall sign of these
effects on the LSD collision rate is therefore unclear.

\begin{figure}
\epsscale{1.1}
\plotone{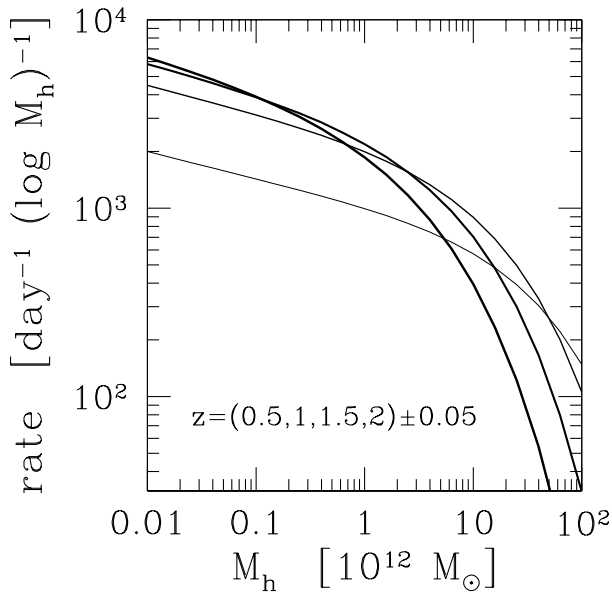}
\vskip .1in
\caption{Differential rate of collisions between LSDs at redshifts $z =$ 0.5, 1, 1.5, 2
(increasing line thickness) in the same calculation shown in Figure \ref{fig:rate}.}
\vskip .2in
\label{fig:rate_halo}
\end{figure}

The LSD velocity distribution function is approximated as locally isothermal and isotropic,
$f_{\rm LSD}(v) = f_0 e^{-v^2/2\sigma^2}$.  The collision rate per unit volume is then
obtained from the cross section (\ref{eq:bcol}) as
\ba\label{eq:rvol}
{d{\cal R}\over dV} &=& \int d^3v_1 \int d^3v_2 \, |{\bf v}_1-{\bf v}_2| \,\sigma(|{\bf v}_1 - {\bf v}_2|)
   \, f(v_1) f(v_2)\nn
     &=&   {2^{2/3}\over \pi^{1/2}}\Gamma\left({4\over 3}\right)\; n_{\rm LSD}^2\sigma \cdot \pi b_{\rm col}^2(\sigma).
\ea
The velocity dispersion profile of each halo is obtained by integrating the Jeans equation \citep{bt08}
\ba
{r_s\over \rho_d}{d(\rho_d\sigma^2)\over dr} 
&=& -{GM_h(<r)\over r_s(r/r_s)^2} \nn
&=& -{3\over 2}(Hr_s)^2 {\delta_c\over (r/r_s)^2}\times \nn
&&\quad \left[\ln\left(1+{r\over r_s}\right) - {r/r_s\over 1+r/r_s}\right].\quad\null
\ea

Finally the observed rate of potentially detectable LSD collisions is
\be
{d\Gamma_{\rm col}\over dz} 
    =  {4\pi d_L(z)^2 \over (1+z)^3} {c\over H(z)} \left\langle {d{\cal R}\over dV}\right\rangle(z).
\ee
Here a factor $(1+z)^{-1}$ accounts for cosmological time dilation, and the luminosity distance is $d_L(z)  
= (1+z) c\int dz/H(z)$.  The average
collision rate density $\langle d{\cal R}/dV\rangle$ at redshift $z$ is obtained by integrating the 
volumetric rate (\ref{eq:rvol}) over the halo in each mass bin, and then summing over the halo mass 
function.

The cumulative distribution of collision rate with redshift is shown in Figure \ref{fig:rate_halo} 
for LSDs comprising the entirety of the dark matter ($f_{\rm LSD} = 1$), and with mass ${\cal M} = 10^{20}$ g, 
radius ${\cal R} = 0.1$ cm, and electromagnetic inertia parameter $f_{\rm em} = 0.025$ (as appropriate
to `spring' loops:  Section \ref{s:scs}).  Here we include the contribution from all halos with masses
exceeding $10^{10}\,M_\odot$.  The result agrees with the preceding estimate (\ref{eq:colrate}).
The differential contribution of halos of different masses at $z=1$ is shown in Figure \ref{fig:rate_halo}.

A more general dependence of $\Gamma_{\rm col}$ on dipole mass, abundance, and size is obtained from the scaling 
\be\label{eq:colrateb}
\Gamma_{\rm col} \propto \left({f_{\rm LSD}\over {\cal M}}\right)^2 R_{\rm col}^2.
\ee
Changes in ${\cal M}$ still yield a fixed rate $\Gamma_{\rm col}$ if we also vary $\mu$ according to 
\be\label{eq:mumrel}
\mu_{18}({\cal M},f_{\rm LSD},\Gamma_{{\rm col},z<1}) 
= 1.5\,{{\cal M}_{20}^2\over f_{\rm LSD}^{3/2}}\left({\Gamma_{{\rm col},z<1}\over 10^4~{\rm d}^{-1}}\right)^{3/4}\;
{\rm G\,cm^3}.
\ee
Of course variations in detectability with explosion energy will modify this scaling.

When the LSD is composed of superconducting `spring' loops, as described in Section \ref{s:scs}, 
it turns out that the collision rate depends mainly on $f_{\rm LSD}$ and the string 
tension, not separately on ${\cal M}$ or $\mu$.   

Let us consider in more detail the dependence of $\Gamma_{\rm col}$ on the mass and
energy density of the LSDs, or equivalently on the energy scale at which they formed in
the early universe.  Although LSD masses as light as ${\cal M} \sim 10^{18}$ g
are consistent with FRBs of energy $\sim 10^{39}$ erg, the radiative efficiency is 
only modest, $\sim (2\pi \nu {\cal R}/c)^{0.5-1}$ at frequency $\nu$ (Paper I).
The mass and radius normalizations we have chosen correspond, through Equation (\ref{eq:dipmass}), 
to a magnetic field $\sim 10^{22} f_{\rm em}^{1/2} {\cal M}_{20}^{1/2}{\cal R}_{-1}^{-3/2}$ G, 
and an energy density $E_{\rm LSD}^4/(\hbar c)^3 \sim (20~{\rm Gev})^4/(\hbar c)^3$.   Expressing 
$\mu$ in terms of $E_{\rm LSD}$ and ${\cal M}$ shows that
\be
\Gamma_{\rm col} \propto f_{\rm LSD}^2 {\cal M}^{-4/3} E_{\rm LSD}^{-8/3}.
\ee
Not surprisingly, higher ${\cal M}$ and $E_{\rm LSD}$ correspond to lower collision rates.

\section{LSD as a Superconducting `Spring' Loop}\label{s:scs}

The preceding considerations are illustrated here with a concrete example:  
loops of superconducting cosmic string carrying an electric current high enough
to cancel off the string tension.  Superconductivity in topologically stable, relativistic string defects
was demonstrated by \cite{witten85}, under varying assumptions about the coupling of
the carriers of ordinary electric charge to a Higgs field that breaks an additional 
$U(1)$ gauge symmetry.  As the current flowing 
along the string grows in strength, the magnetic field winding around the string, and the kinetic
energy of the charge carriers, together provide a rising positive pressure.  It was conjectured by 
\cite{otw86}, and demonstrated in detail by  \cite{copeland87}, \cite{haws88} and \cite{davis89},
that closed loops of the string could form a stable, macroscopic form of dark matter
due to a cancellation between the string tension and this positive pressure.   If formed at sufficiently
high redshift (associated, e.g., with symmetry breaking at a GUT energy scale
$M_Gc^2 \sim 10^{15}$-$10^{16}$ GeV), these superconducting `springs' could easily overclose the universe.

In fact, macroscopic superconducting `spring' loops with a GUT-scale tension ${\cal T}
\sim 10^{-8}c^4/G$ have just the relation between size and
magnetic moment that we are looking for:  a loop of radius ${\cal R}$ has a mass
\be
{\cal M} \simeq {2\pi {\cal R} {\cal T}\over c^2} = 8.5\times 10^{19}\left({GT\over c^4}\right)_{-8} {\cal R}_{-1}\quad {\rm g}.
\ee
The magnetic field scales inversely with distance 
$\varpi$ from the string as $B_\phi(\varpi) = 2I/c\varpi$, meaning that a relatively weak field
$2\mu/{\cal R}^3 \sim 10^{22}$ G is sustained at a macroscopic scale ${\cal R} \sim 0.1$ cm
-- as compared with the more substantial field ($\sim 10^{48-50}$ G) encountered in the string core.
The one-dimensional pressure imparted by this field, integrated from the core to radius ${\cal R}$, is
\be
{dE_B\over d\ell} = \int_{R_G}^{{\cal R}} {B_\phi^2(\varpi)\over 8\pi} 2\pi\varpi d\varpi = 
      {I^2\over c^2} \Lambda 
\ee
Here $R_G \equiv \hbar/M_G c$ and $\Lambda = \ln({\cal R}/R_G) \sim 60-70$.  

This large logarithmic factor greatly enhances the inertia of the loop, relative to the energy
stored in the magnetic field at a scale $\sim {\cal R}$, which is approximately
the magnetic dipole power $\sim \mu^2 {\cal R}^{-3}$ that would be radiated during a collision between
similarly sized loops.  Adopting the parameterization (\ref{eq:dipmass}), and noting that a circular
loop carrying current $I$ has a magnetic moment $\mu = \pi {\cal R}^2 I/c$, one finds
\be\label{eq:fem}
f_{\rm em} = {\pi \over 2\Lambda} \simeq 0.025.
\ee

The relation between loop mass and radiated electromagnetic energy $d{\cal E}_{\rm em}/d\ln\nu$
in a given band is obtained as follows.  The initial pulse width is $\sim {\cal R}$ and the radiative 
efficiency at low frequency $\nu$ is $\sim (2\pi \nu {\cal R}/c)^{0.5-1}$ (that is, at a wavelength much larger 
than $\sim 2\pi {\cal R}$; Paper I).   So, for example, for an index 0.5,
\ba\label{eq:Eem}
{d{\cal E}_{\rm em}\over d\ln\nu} &\sim& \left({2\pi \nu {\cal R}\over c}\right)^{1/2} f_{\rm em} {\cal M}c^2 \nn
&=& 1\times 10^{40}\,f_{\rm em}{{\cal M}_{20}^{3/2}\nu_9^{1/2}\over [(G{\cal T}/c^4)_{-8}]^{1/2}}\quad {\rm erg}.
\ea

There is an additional subtlety here arising from the sign of the 
current flowing along the string, relative to the longitudinal magnetic flux quantum threading 
its core.  (This flux is associated with $U(1)$ gauge symmetry whose spontaneous breaking
gives rise to the string defect.)  Two gauged string loops of the same type always reconnect so that the direction
of the core flux is continuous across the reconnection point.  If their currents have the same relative
sign, then it is easy to see that Equation (\ref{eq:fem}) gives a good estimate of the
ratio of radiated energy to rest energy of the colliding loops.  Consider two `spring' loops
of equal sizes ${\cal R}$ and (necessarily) equal currents.   The magnetic energy of each
loop before the collision is, in a first approximation, $E_B = 2\pi {\cal R} (I/c)^2 \ln({\cal R}/R_G)$.
After the collision, the energy is $E_B = 2\pi (2 {\cal R} + \delta {\cal R}) (I/c)^2 \ln(2{\cal R}/R_G)$. 
Even in the absence of radiative energy loss during the collision, a small shrinkage 
$\delta {\cal R} \sim -{\cal R}/\Lambda$ is required to restore vanishing net tension.

The dissipation is much greater when the colliding `spring' loops have currents with opposing
signs.  Now when the strings reconnect, the current develops strong inhomogeneities
and zones with finite linear charge density $\rho$, even exceeding $|I|/c$ in magnitude in places.  Here the electromagnetic field surrounding the string is predominantly electric, and one
expects very strong dissipation.  Collisions between loops with opposing currents should
therefore produce explosions with a low-frequency electromagnetic precursor of energy
similar to that radiated in the case of aligned currents, but followed by an intense 
thermal fireball.  Now $f_{\rm em}$ represents the ratio of energies in these two components.   
The fireball can itself couple effectively to a propagating electromagnetic mode in the surrounding
plasma (\citealt{blandford77}, Paper I).

Next we consider the collision rate of superconducting `spring' loops as a function of
the string tension.  Recall that the collision radius $R_{\rm col}$ is proportional to the radius
${\cal R}$ of the dipole but does not depend explicitly on its mass.  Equation (\ref{eq:colrateb}) then
shows that the collision rate scales as $\Gamma_{\rm col} \propto f_{\rm LSD}^2({\cal R}/{\cal M})^2$.
But the string tension is approximately ${\cal T} \sim {\cal M}c^2 / 2\pi {\cal R}$, and so
we find that $\Gamma_{\rm col} \propto (f_{\rm LSD}/{\cal T})^2$.  The normalization works out to
\be
\Gamma_{{\rm col},z<1} = 1\times 10^4\,f_{\rm LSD}^2\left({G{\cal T}/c^4\over 10^{-8}}\right)^{-2}\quad {\rm day}^{-1}.
\ee
The rate as a broader function of redshift is shown in Figure \ref{fig:rate} for 
$f_{\rm LSD}^{-1}(G{\cal T}/c^4) = 10^{-8}$.

Our ab inito rate calculation can be compared with the measurement $\Gamma_{\rm FRB} = 3\times 10^3$ d$^{-1}$
for FRBs with fluence larger than $F_{\nu,\rm min} = 3.8$ Jy-ms at a frequency $\nu \sim 1.4$ GHz \citep{crawford16}.  
The curve in Figure \ref{fig:rate} matches $\Gamma_{\rm FRB}$ at a cumulative
emission redshift $z_{\rm max} \sim 0.55$.  The rest-frame radio energy implied by a fluence $F_{\nu,\rm min}$ from a
redshift $z_{\rm max}$ is  $E_{\rm FRB} = 4\pi (1+z_{\rm max})^{-1} d_L(z_{\rm max})^2 \nu F_{\nu,\rm min} \sim 4\times 10^{40}$
erg.  If this energy were supplied by the collision of two LSDs, then the 
mass implied by by Equation (\ref{eq:Eem}) is ${\cal M} \sim (1-2)\times 10^{20}$ g (again for ${\cal T} = 10^{-8}c^4/G$).

We note in closing that relativistic cusps forming on oscillating cosmic strings have long been known to
be sources of intermittent bursts of gravitational and electromagnetic waves.  In this case, the 
emitted electromagnetic pulse is significantly longer than a radio wavelength, so that the calculations
of deceleration and radio wave emission in Paper I are not applicable.  More broadly emission with a
sub-millisecond duration, or emission repeating on a timescale of tens or hundreds of seconds, 
is not a natural outcome of GUT-scale string dynamics.  A recent exploration in the context of FRBs 
has been made by \cite{yu14}.

\section{Collisional LSD Rings Trapped Near Supermassive Black Holes}\label{s:smbh}

A competing, and perhaps dominant, channel for LSD annihilation involves SMBHs.
This channel is made effective by a remarkable combination of
properties of SMBHs:  i) an enormous cross section to accrete ambient material, especially during a rapid growth phase;
ii) the small volume of the space surrounding the hole, which permits a high optical
depth to collisions between LSDs trapped there; and iii) the presence of gas dense enough to
drag LSDs into a narrow and long-lived ring just outside the ISCO.

\subsection{Adiabatic Compression of LSD Dark Matter around SMBHs Formed by Direct Gas Collapse}

We first estimate the net mass of LSD dark matter that would be compressed close enough to a
SMBH to allow gas drag to take over and bring the LSDs close to the ISCO.   This mass is
maximized if the hole is assembled in situ in a galactic nucleus, in particular if most of its growth
is due to hyper-Eddington accretion from a 10-100 pc sized plasma cloud \citep{loeb94,begelman06}.
Dark matter permeating the nuclear region is adiabatically compressed as the black hole grows, with
the strength of the compression depending on the relative increase in hole mass.
The following considerations are therefore most relevant for the first generation of SMBHs,
which might form in spheroids of mass $\sim 10^9\,M_\odot$ or smaller.  

The dark matter is assumed initially to have formed a constant density core with central
velocity dispersion $\sigma_d$.  Once the SMBH has collapsed, its growth
is filled out by continued radiative accretion, and hierarchical merging of with other
dark matter halos brings the central stellar velocity dispersion up to the value implied by the 
measured $M_\bullet-\sigma$ relation \citep{gultekin09}.  Consistent with this approach, we 
normalize $\sigma_d$ to $\sim 40$ km s$^{-1}$ for a $10^6\,M_\odot$ hole.

The gas profile around the nascent hole during the super-Eddington accretion phase is approximated
by a singular isothermal sphere,
\be\label{eq:rhog2}
\rho_g(r) \sim {\sigma_g^2\over 2\pi G r^2}.
\ee
Here $\sigma_g$ incorporates the contributions of turbulent, magnetic, and rotational stresses
within the contracting gas, as well as strong radiation pressure.  Collectively these stresses
slow down the collapse compared with a thermally supported sphere.  This type of structure is observed,
with a normamlization $\sigma_g \sim 20$ km s$^{-1}$, in the gas$+$N-body simulations of 
low-mass galactic halos performed by \cite{choi13,choi15}.

At the end of the prompt accretion phase, after the hole has developed and most of the gas
around it has dissipated, the dark matter profile is given by \citep{peebles72,young80}
\be\label{eq:rhod0}
{\rho_d(r)\over \rho_{d,0}} = {4\over 3\pi^{1/2}}\left({GM_\bullet\over \sigma_d^2 r}\right)^{3/2}.
\ee
We normalize the initial central dark matter density $\rho_{d,0}$ in terms of a transition radius 
$R_{\rm gas}$ where $\rho_g(R_{\rm gas}) = \rho_{d,0}$.  Then the mass profile of the dark matter
cusp around the SMBH is
\ba
{1\over M_\bullet}{dM_d\over d\ln r_d} &=& \left({r_d\over R_g}\right)^{3/2} 
{8G^2 M_\bullet^2\sigma_g^2\over 3\pi^{1/2}c^3\sigma_d^3 R_{\rm gas}^2}\nn
 &=& 2\times 10^{-9} \left({r_d\over 10^3\,R_g}\right)^{3/2}
{M_{\bullet,7}^2\sigma_{g,20}^2\over \sigma_{d,40}^3 (R_{\rm gas}/10^2~{\rm pc})^2},\nn
\ea
where $\sigma_{g,20} = \sigma_g / (20~{\rm km~s^{-1}})$, $\sigma_{d,40} = \sigma_d / (40~{\rm km~s^{-1}})$.
We see that this process would allow $\sim 10^{12}$ LSD particles each of mass $10^{20}$ g to accumulate
within $10^3\,R_g$ of a $10^7\,M_\odot$ SMBH.   

The hydromagnetic drag experienced by the LSDs produces a strong enhancement in annihilation efficiency
around a SMBH, as compared with dark matter composed of weakly interacting massive particles.   In the latter case,
one must assume that the initial dark matter profile is already centrally peaked (e.g. $\rho_d(r) \propto r^{-1}$:  
\citealt{nfw97}) in order to obtain an interesting WIMP annihilation rate \citep{gondolo99}.   (In this
respect, relatively conservative assumptions underly our estimate of the LSD annihilation rate.)  A sharper
final dark matter cusp is also obtained if the initial orbits of the dark matter particles are circular
\citep{steigman78}, but this density profile is not consistent with conservation of
phase space density.  Simulations by \cite{capela14} assuming a more isotropic initial phase space distribution 
produce a final density profile close to $\rho_d(r) \propto r^{-3/2}$, in agreement with the classic stellar dynamical
result.

\begin{figure}
\epsscale{1.1}
\plotone{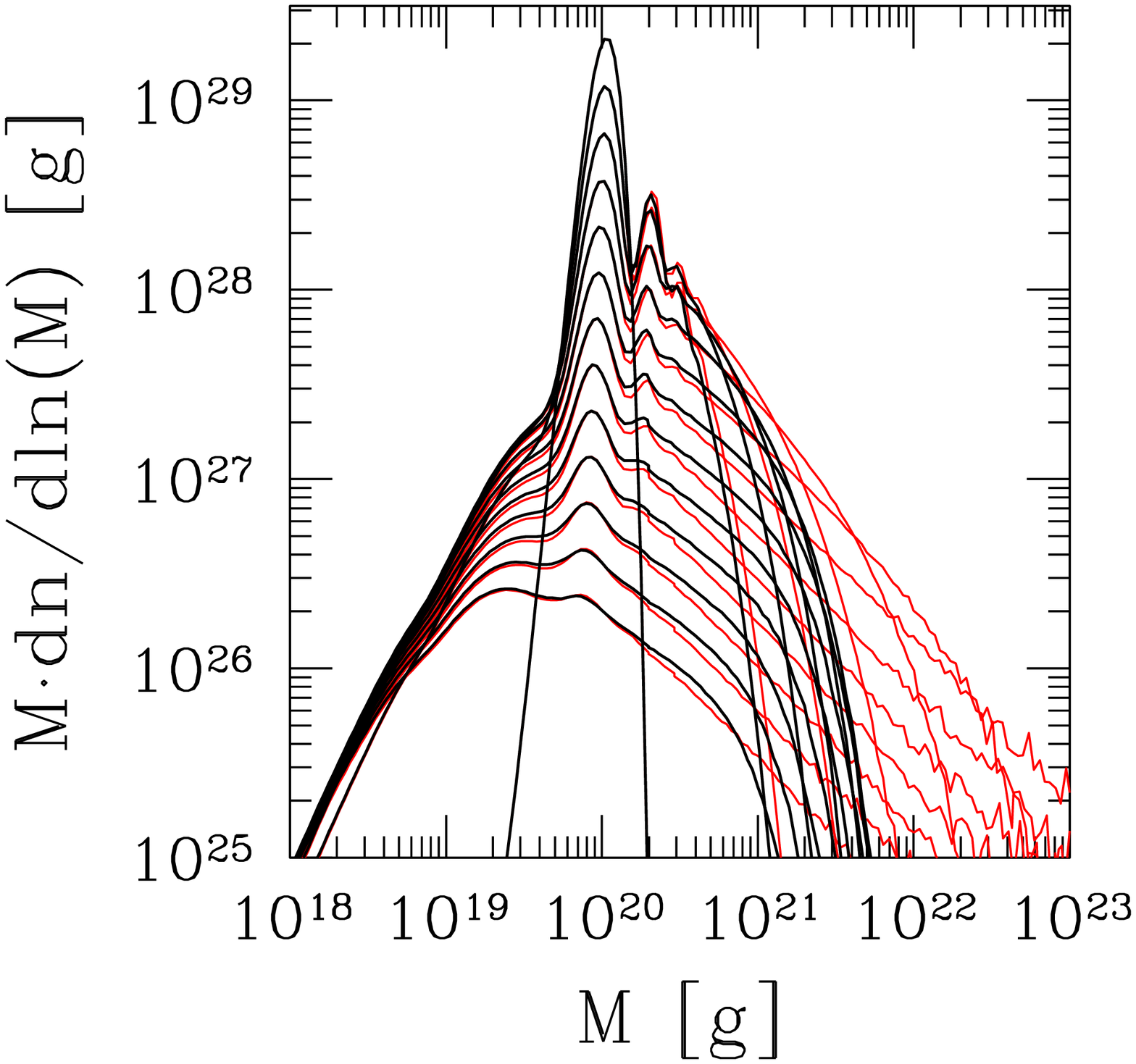}
\plotone{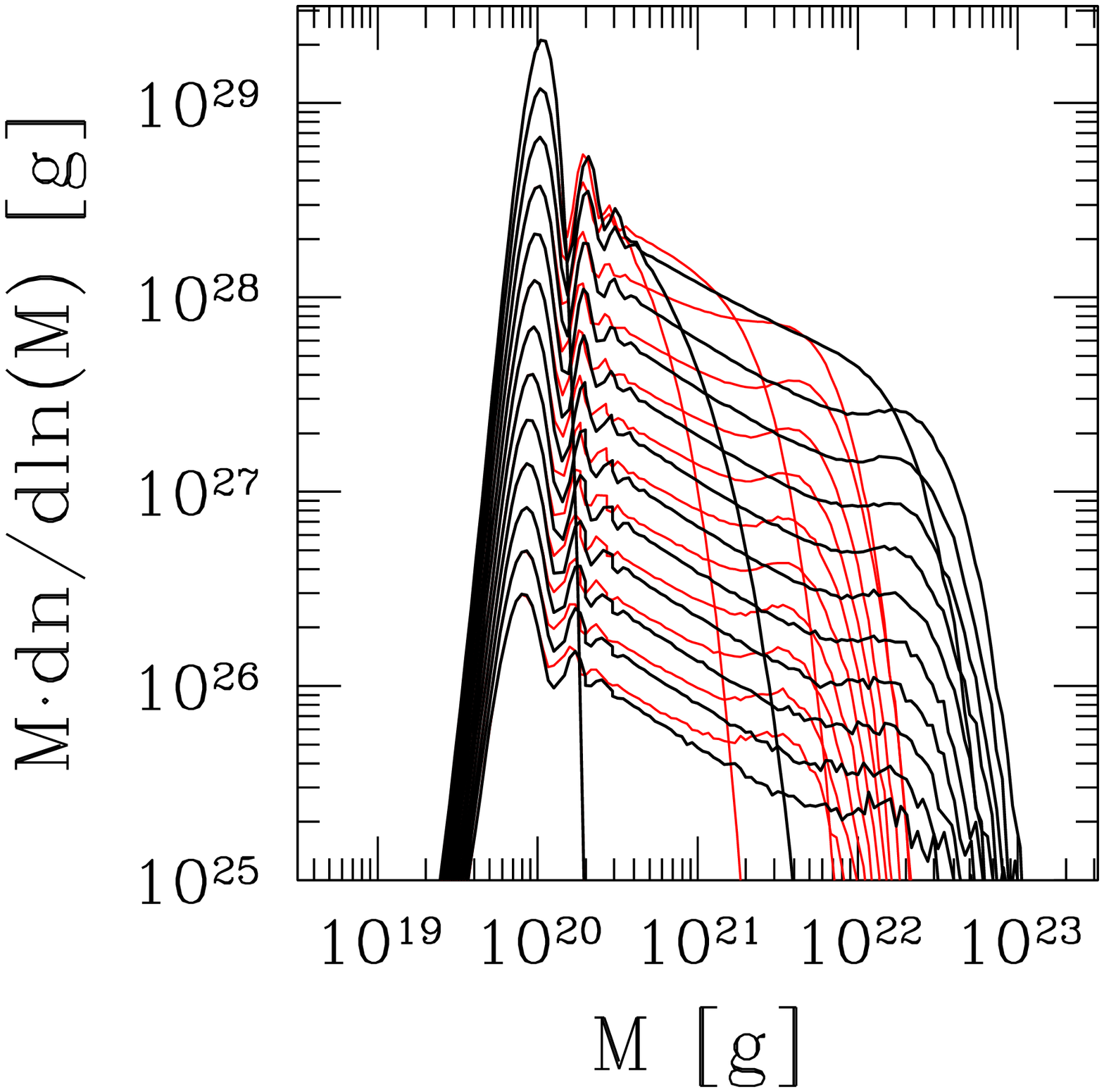}
\caption{Collisional evolution of $10^9$ dipoles trapped in a small volume, obtained from a direct Monte Carlo 
integration.  Gaussian initial mass spectrum centered around
$\bar{\cal M} = 10^{20}$ g.     Cross section for collision between two dipoles of 
masses ${\cal M}_1 \neq {\cal M}_2$ and $\mu_1 \neq \mu_2$ is given by Equation (\ref{eq:cs12}).  We treat separately two
cases, appropriate to superconducting `spring' loops, in which each loop has a random sign of $I\cdot \Phi_s$
({\it top panel}) or ii) a fixed sign of $I\cdot \Phi_s$ ({\it bottom panel}).  (Here $I$ is the electric current flowingalong the string
and $\Phi_s$ is the magnetic flux quantum (associated with a spontaneously broken $U(1)$ gauge symmetry) 
threading the string core.)
Different line colors show the effect of varying the fraction $f_{\rm em}$ of the energy radiated in a given collision.
Black curve:  $f_{\rm em} = 0.025$ when the colliding loops have the same sign of $I\cdot \Phi_s$;  otherwise
(only in case i)) $f_{\rm em} = |{\cal M}_1 - {\cal M}_2|/({\cal M}_1 + {\cal M}_2)$.
Red curve:  $f_{\rm em} \rightarrow 0$ ({\it top panel});  $f_{\rm em} \rightarrow 0.1$ ({\it bottom panel}).}
\vskip .2in
\label{fig:mass_spectrum}
\end{figure}

\begin{figure}
\epsscale{1.1}
\plotone{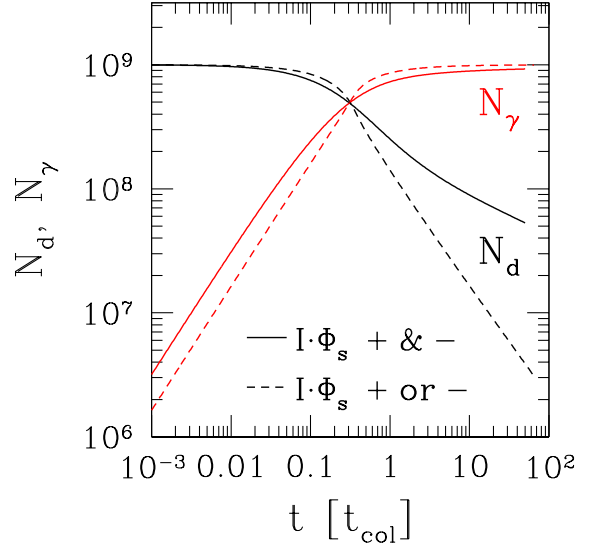}
\caption{Rate of change of total dipole number $N_d$ (black curves) and cumulative number of electromagnetic bursts
$N_\gamma$ (red curves) in the simulations of Figure \ref{fig:mass_spectrum}.   Time is scaled to collision time of
two dipoles each of mass $\bar {\cal M}$ using Equation (\ref{eq:colrate2}).  Solid curves:  
case i) in which loops have both signs of $I\cdot \Phi_s$ with equal probability.  Dashed curves:  case ii)
in which $I\cdot \Phi_s$ takes only one sign.  Baseline $f_{\rm em} = 0.025$ in both cases.  The decay of $N_d$
is slower in case i) because the net dipole mass, and the collision cross section (\ref{eq:cs12}),
drops in larger steps when loops of opposing signs collide.}
\vskip .2in
\label{fig:time_profile}
\end{figure}

\begin{figure}
\epsscale{1.1}
\plotone{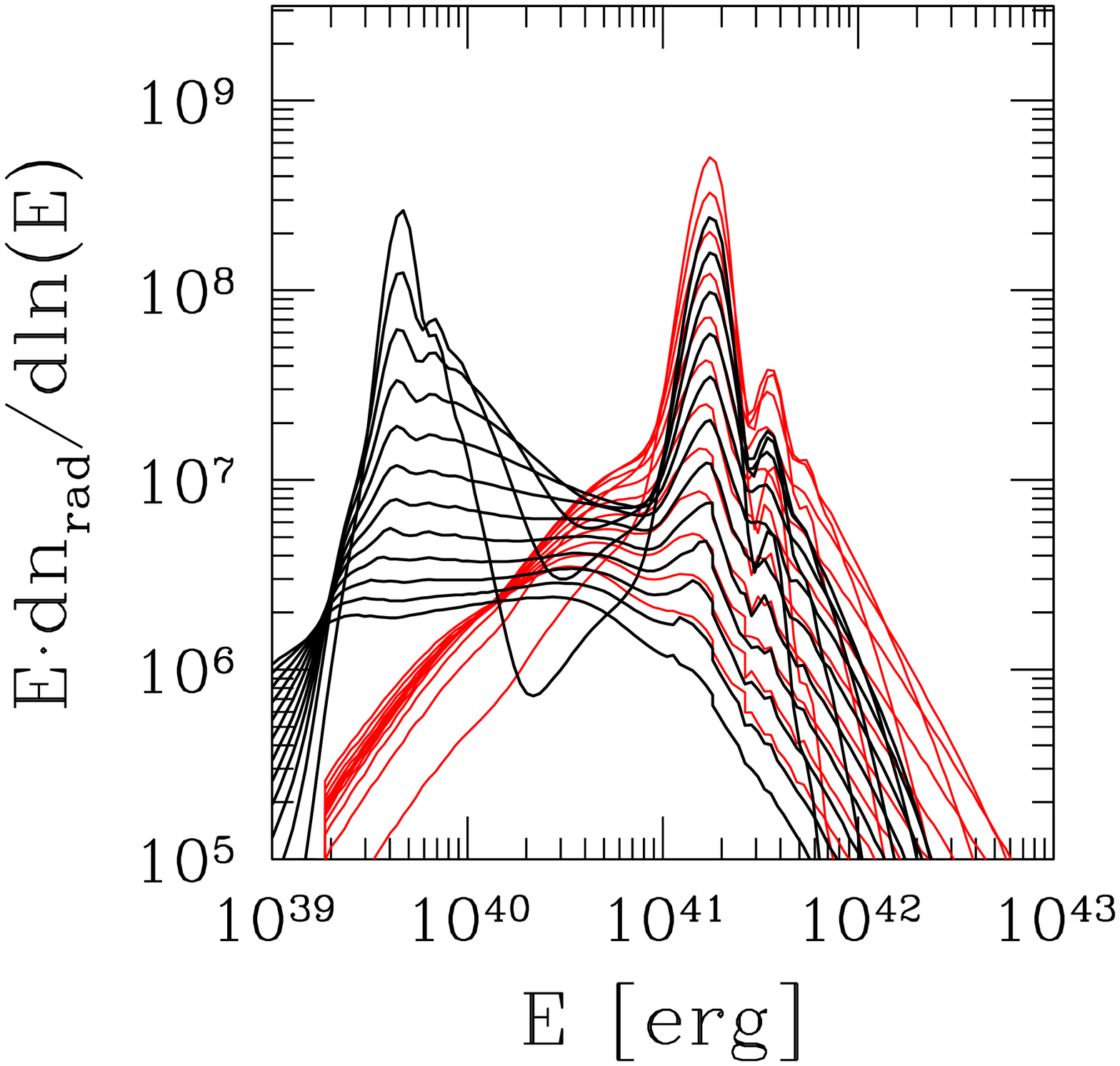}
\plotone{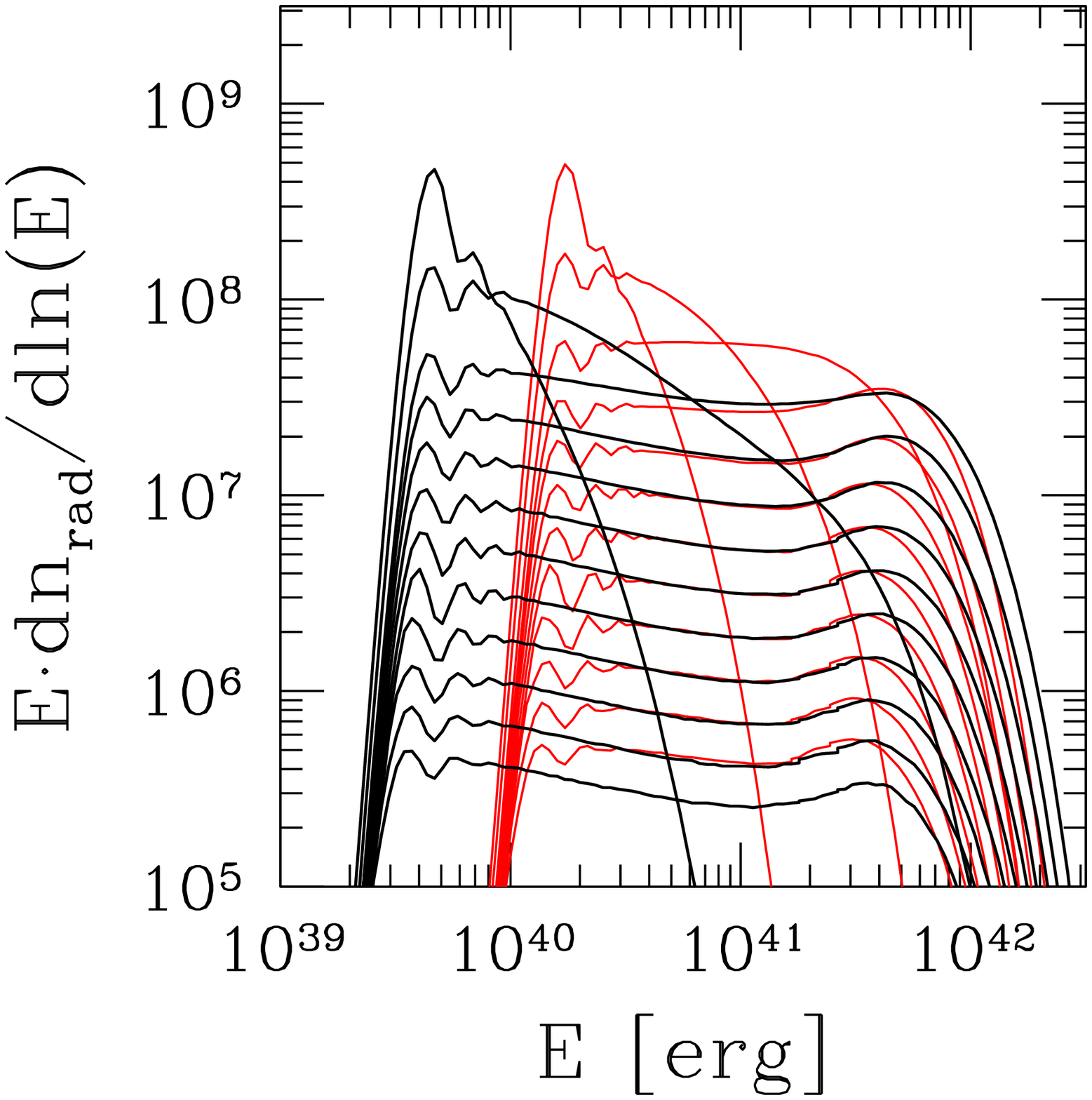}
\caption{Distribution of bolometric energies of low-frequency electromagnetic bursts produced by collisions between dipoles
in the simulations shown in Figure \ref{fig:mass_spectrum}.   {\it Top panel:}  case i) with $I\cdot \Phi_s$
having random sign.  {\it Bottom panel:}  case ii) in which $I\cdot \Phi_s$ has uniform sign.  Black curves:
$f_{\rm em} = 0.025$.  Red curves:  $f_{\rm em} \rightarrow 0$ ({\it top panel}) and $f_{\rm em} = 0.1$ ({\it bottompanel}).}
\vskip .2in
\label{fig:energy_spectrum}
\end{figure}

\subsection{Collisional Evolution of a LSD Ring \\ Just Outside the ISCO}

Consider a large number $N_d$ of LSDs that orbit in the equatorial plane of the black hole,
near the radius where $dP_g/dr = 0$ in the accreting plasma flow.   We ignore relativistic effects here and limit ourselves
to a Newtonian treatment of the gravitational interactions of the dipoles.   In this environment,
gravitational stirring can be an important agent for growing random velocities.  Expression (\ref{eq:vmax})
approximates the maximum random speed that can be reached by gravitational dipole-dipole scattering
before it is limited by dissipative collisions.  

The tidal force exerted by the SMBH has a negligible effect on LSD
interactions when $v \gtrsim \Omega a ({\cal M}/M_\bullet)^{1/3} \sim 10^{-7} c\,{\cal M}_{20}^{1/3} M_{\bullet,7}^{-1/3}$.
Then the scattering impact parameter of two dipoles each of mass ${\cal M}$ is $\sim 2G{\cal M}/v^2$.
Balancing this with the collision radius (\ref{eq:rcol}) gives Equation (\ref{eq:vmax}).  

We suppose that the dipoles placed near the ISCO initially feel strong enough drag that their orbits nearly
circularize before two-body interactions become important.  After this, the gas accretion rate is
reduced, and the dipoles begin to interact gravitationally with each other.   Their random velocity grows by gravitational
scattering until it reaches the value (\ref{eq:vmax}), after which the number and mass of the dipoles
is reduced by collisions mediated by the magnetic interaction.  

The effective volume of the ring occupied by the LSDs 
is $V_{\rm ring} \sim 2\pi a \cdot \pi h^2 = \pi^3 a (v/\Omega)^2$.   Approximating
collisions as complete annihilations, the corresponding rate is
\be\label{eq:colrate2}
\dot N_d \sim -{N_d^2\over V_{\rm ring}}\,\sigma_{\rm col} v.
\ee
Since $v$ is independent of $N_d$ but $\sigma_{\rm col} v \propto v^{-1/3}$ (Equation (\ref{eq:bcol})), 
we obtain the scaling solution
\ba\label{eq:nd}
N_d(t) &\sim& {\pi^2\over 3f_{\rm em}^{2/3}}\left({v\over c}\right)^{7/3}\left({a\over R_g}\right)^4 
                  \left({R_g\over {\cal R}}\right)^2 {R_g\over ct}\nn
     &=& 3\times 10^5\,M_{\bullet,7}^3 {\cal M}_{20}^{7/4} {\cal R}_{-1}^{-15/4}
         \left({t\over {\rm Gyr}}\right)^{-1}
\ea
at $a \sim 6R_g$ and for $f_{\rm em} = 0.025$.  

The steadily decreasing dipole number (\ref{eq:nd}) is enhanced by a steady inflow of dipoles,
say at a rate $\dot N_{d+}$.  Balancing this source with annihilation gives the equilibrium number
\be\label{eq:ndeq}
N_{d,\rm eq} = 6\times 10^7\, M_{\bullet,7}^{3/2} 
{\cal M}_{20}^{7/8}{\cal R}_{-1}^{-15/8}\left({\dot N_{d+}\over 10\,{\rm yr}^{-1}}\right)^{1/2}.
\ee
This result only holds at a sufficiently advanced time.

\subsection{Collisional Evolution of the Dipole Mass Spectrum}

We now turn to the effects of collisions on the mass spectrum, and the energy distribution
of radiated electromagnetic bursts that results.   Start with a narrow initial distribution
of ${\cal M}$ centered around some  mass $\bar{\cal M}$ (Figure \ref{fig:mass_spectrum}),
with a corresponding magnetic moment $\bar\mu$.  We
focus on the specific case of superconducting `spring' loops.  These can exist in two
configurations with $I\cdot\Phi_s > 0$ or $<0$, where $I$ is the electric current flowing 
along the string, and $\Phi_s$ is the magnetic flux quantum threading the string core.  

The cross section for two dipoles with $\mu_1 \neq \mu_2$ is obtained by combining
Equations (\ref{eq:rcolb}) and (\ref{eq:bcol}), and assuming proportionality of $\mu$ and ${\cal M}$, as
for a `spring' of fixed tension ${\cal T}$.  
\ba\label{eq:cs12}
\sigma_{\rm col}(\mu_1,\mu_2) &=& \sigma_{\rm col}(\bar\mu,\bar\mu) {(\mu_1/\bar\mu)^{1/3}\,
(\mu_2/\bar\mu)^{1/3} \over (2{\cal M}_r/\bar {\cal M})^{1/3}}\nn
&= & \sigma_{\rm col}(\bar\mu,\bar\mu) \, \left({{\cal M}_1 + {\cal M}_2\over 2\bar {\cal M}}\right)^{1/3}.
\ea
Here ${\cal M}_r = {\cal M}_1{\cal M}_2 / ({\cal M}_1 + {\cal M}_2)$ is the reduced mass. 

Here we ignore the absolute normalization of time and dipole density.  We start with a 
large number ($\sim 10^9$) of dipoles, evolving their mass distributions over a multiple of
the collision time as defined by the initial configuration.
The top panel
of Figure \ref{fig:mass_spectrum} shows the result when equal numbers of both signs of
$I\cdot \Phi_s$ are present
in the initial condition.  Collisions between loops of equal sign of $I\cdot \Phi_s$ create 
a loop whose mass is only slightly smaller than the sum, by a fractional amount $f_{\rm em} = 0.025$.  
The red curves show the effect of removing this small radiative loss ($f_{\rm em} \rightarrow 0$), but still 
allowing for the much stronger dissipation that results when loops of opposing signs collide.  
An extended power-law tail forms at high masses, which is truncated at a mass $\sim f_{\rm em}^{-1}
\bar{\cal M}$ for finite $f_{\rm em}$.

A different result is obtained when only loops with one sign of $I\cdot \Phi_s$ are present
(bottom panel of Figure \ref{fig:mass_spectrum}).  Now
strongly dissipative mergers are absent, and finite $f_{\rm em}$ is required to prevent most
of the mass from collecting in very large loops.  As before, increasing $f_{\rm em}$ reduces
the mass at which the high-energy tail cuts off.

Figure \ref{fig:time_profile} shows how the number of loops decreases with time in the two cases.
The decrease is slower when loops of both signs are present, because more small
loops are formed which interact with a reduced cross section (\ref{eq:cs12}).  This effect
is evident in the distribution of electromagnetic pulse energies (Figure \ref{fig:energy_spectrum}), 
which shows a main peak at an energy $\sim \bar{\cal M} c^2$ and a separate, lower-energy peak at 
$f_{\rm em} \bar{\cal M}c^2$.  A much broader high-energy pulse energy tail is present when only loops
of a single sign of $I\cdot \Phi_s$ are present (lower panel of Figure \ref{fig:energy_spectrum}).

\section{Collisions with White Dwarf Stars}\label{s:wd}

Stars of all types can collide with a LSD orbiting through the same galaxy.  
In contrast with small primordial black holes (PBHs), whose interactions with stars
have been considered by \cite{capela13} and \cite{graham15}, a trapped LSD cannot destroy its host by
accreting it.   But high
temperatures are generated behind the shock that forms around the LSD during its supersonic motion through the 
star.   In the case of a WD, one must consider whether this heating will trigger runaway thermonuclear burning.   
\cite{graham15} considered the interaction of small PBHs with WDs, finding a minimal size for the PBH that would
trigger a carbon deflagation in WDs of various masses.   
The contribution of PBHs of mass $10^{20}$-$10^{22}$ g to the cosmic
dark matter was found to be constrained by the observed rate of Type Ia supernovae.

Here we show that LSDs, of the mass and size needed to explain FRBs, will trigger thermonuclear deflagrations,
and possibly direct detonations, in C/O WDs with masses exceeding $\sim 1\,M_\odot$.   
The collision probability is high for a star of mass $\sim M_\odot$ to experience such a collision
if its lifetime exceeds a billion years, and if $f_{\rm LSD} = O(1)$.  Heavier WDs with ONeMg composition have
a somewhat higher threshold for ignition due to the stronger coulomb barrier in oxygen burning, with the
result that only stars heavier than $\sim 1.29\,M_\odot$ are found to experience self-sustained burning.  
The initial spark is placed well off the center of the star in all cases, but especially in the more massive ONeMg models.
 
We first show that the hydromagnetic drag force overcomes gravity as the LSD moves through a dense and
degenerate star (a WD or neutron star).
Then the differential speed $\Delta v$ between LSD and stellar material drops below the escape speed, except
for trajectories that only touch the outermost layers.

We begin by evaluating the drag force (\ref{eq:fdrag}) for a LSD on a ballistic trajectory midway through a 
target star of mass $M_\star$ and radius $R_\star$ (e.g. at a radius $\sim R_\star/2$ and local density 
$\rho_{\rm ex} \sim 3M_\star/4\pi R_\star^3$).  The magnetic moment is expressed in terms of the string tension ${\cal T}$ 
(for a cosmic `spring' loop), giving
\ba
{F_{\rm drag}\over G{\cal M}M_\star/r^2} &\sim& 0.03\,C_d\left({G^2{\cal M}_{\rm rad}M_\star\over R_\star^2 c^4}\right)^{1/3}
\left({G{\cal T}\over c^4}\right)^{-1}\nn
   &=& 0.02\, {C_d {\cal M}_{\rm rad,20}^{1/3}\over (G{\cal T}/c^4)_{-8}} \left({M_\star\over M_\odot}\right)^{1/3}
\left({R_\star\over R_\odot}\right)^{-2/3}.\nn
\ea
One sees that a solar-type star may or may not supply enough drag to trap a LSD whose differential speed at
infinity is $v_{\infty} \sim 0.3v_{\rm esc}(R_\star) \sim 200$ km s$^{-1}$,
where $v_{\rm esc}(R_\star) = (2GM_\star/R_\star)^{1/2}$ is the escape speed
from the surface of the star.   But the LSD will be trapped by a more compact star.

Consider in particular the interaction of a LSD with a WD composed mainly of carbon and oxygen.  The rate
of carbon burning is sensitive to temperature \citep{caughlan88}, and therefore to details of hydrodynamic
response of the WD material.   In contrast with an infalling PBH, the disturbance
has a large enough scale that conductive heat transport is negligible.   But we show that adiabatic cooling
can have a dramatic effect:  the burning rate drops precipitously as the shocked stellar material expands and returns to
the ambient stellar pressure.  The velocity shear established by the propagating shock can further suppress
thermonuclear burning by triggering turbulence that mixes heat away from the center-of-mass trajectory of the LSD.

First we establish that the interaction between the LSD and the star is mainly hydromagnetic, not gravitational.  
In spite of the enormous difference in material density and magnetic field strength,
there are similarities with the interaction between a planetary magnetosphere and the Solar wind.
The gravity of the LSD by itself deflects the WD material only through a small angle
\be
{\Delta v_\perp\over \Delta v} \sim {G{\cal M}\over R_{\rm mag} \Delta v} \sim 2\pi \left({G{\cal T}\over c^4}\right)
\left({\Delta v\over c}\right)^{-7/3}\left({\bar\rho_{\rm LSD}\over\rho_{\rm wd}}\right)^{-1/6},
\ee
which works out to $\Delta v_\perp/\Delta v \sim 10^{-3} (G{\cal T}/c^4)_{-8}$ for a LSD velocity 
$\Delta v \sim 10^3$ km s$^{-1}$.  Here $\bar\rho_{\rm LSD}$ is given by Equation (\ref{eq:barrho}) and 
$\Delta v_\perp$ is the transverse impulse.

Peak temperature is reached behind the shock close
to the center-of-mass trajectory of the LSD.  The first thing to note is that this temperature only depends on
the mass profile and gravity of the WD, as long as the LSD moves on a nearly ballistic trajectory.  
From Equation (\ref{eq:tdrag}), we see that a LSD `spring' with tension ${\cal T} \sim 10^{-8}c^4/G$
will experience drag inside a WD that is only moderately stronger than gravity.  
The internal energy density near the center of the WD increases rapidly with stellar mass, meaning that there
is a critical mass above which self-sustained carbon (and oxygen) burning is triggered.

\begin{figure}
\epsscale{1.1}
\plotone{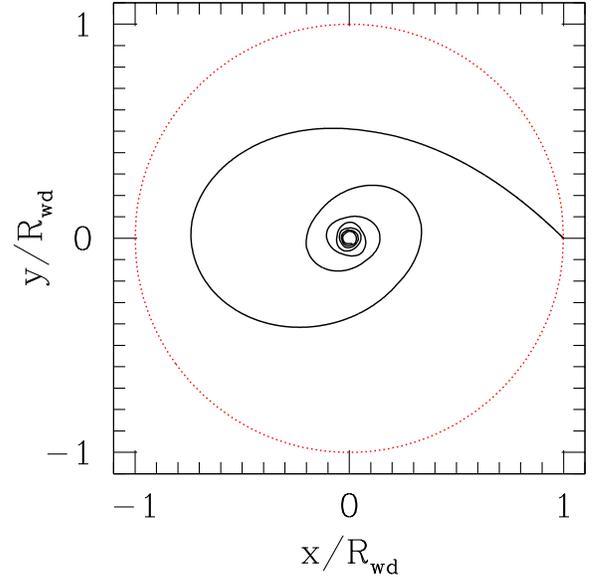}
\vskip .1in
\caption{Sample trajectory of LSD `spring' of mass $4\times10^{20}$ g and tension ${\cal T} = 10^{-8}c^4/G$ falling throughaC/O 
WD of mass $1.02\,M_\odot$ and central density $10^{7.5}$ g cm$^{-3}$, in response to the stellar gravity and 
the hydromagnetic drag force (\ref{eq:fdrag}).   LSD velocity at infinity $\Delta v_\infty = 200$ km s$^{-1}$ and 
impact parameter $b = b_{\rm col}/\sqrt{2}$.  White dwarf surface is marked by the dotted red circle.}
\vskip .2in
\label{fig:trajectory}
\end{figure}

We proceed by introducing LSDs on downward Keplerian trajectories at the surface of a model WD, and then
follow their radial and non-radial motion inside the star in response to the central gravity and drag force 
(\ref{eq:fdrag}).  The initial orbital angular momentum (impact parameter $b$) is scaled to the maximum value 
$\ell_{\rm col} = (2GM_{\rm wd} R_{\rm wd})^{1/2} $ that allows a direct collision with a star of mass 
$M_{\rm wd}$ and radius $R_{\rm wd}$,
\be
\ell = b \Delta v_\infty = \left({b \over b_{\rm col}}\right) \ell_{\rm col}; \quad\;
b_{\rm col} \equiv {(2 GM_{\rm wd} R_{\rm wd})^{1/2}\over \Delta v_\infty}.
\ee
The trajectory is characterized by $\ell$, energy ${\cal E}$, and radial speed $\Delta v_r$.  The energy
is initialized to ${\cal E} = {1\over 2}\Delta v_\infty^2$, but the burning behavior is not sensitive to
this choice for values of $\Delta v_\infty$ characteristic of a galactic potential, being dominated instead
by the stellar gravity.   These quantities evolve according to
\ba
{dr\over dt} &=& \Delta v_r = \pm 2^{1/2}\left[{\cal E} - \Phi(r) - {\ell^2\over 2r^2}\right]^{1/2};\nn
\Delta v &=& 2^{1/2}\left[{\cal E} - \Phi(r)\right]^{1/2};\quad {d\Phi\over dr} = {GM_{\rm wd}(<r)\over r^2};\nn
{d{\cal E}\over dt} &=& - F_{\rm drag}\Delta v; \quad\quad
{d\ell\over dt} = -F_{\rm drag}{\ell\over \Delta v}.
\ea
Here $M_{\rm wd}(<r) < M_{\rm wd}$ is the mass enclosed inside radius $r < R_{\rm wd}$.
The drag force evolves according to Equation (\ref{eq:fdrag}) in response to changes in $\Delta v$
and ambient mass density $\rho_{\rm ex} = \rho_{\rm wd}(r)$.  

A sample trajectory of an infalling LSD particle (a superconducting `spring' of mass $10^{20}$ g and tension
${\cal T} = 10^{-8}c^4/G$) spiraling through a $1.02\,M_\odot$ C/O WD is shown in Figure \ref{fig:trajectory}.
The LSD impact parameter is $b_{\rm col}/\sqrt{2}$ and velocity at infinity $v_\infty = 200$ km s$^{-1}$.

\begin{figure}
\epsscale{1.1}
\plotone{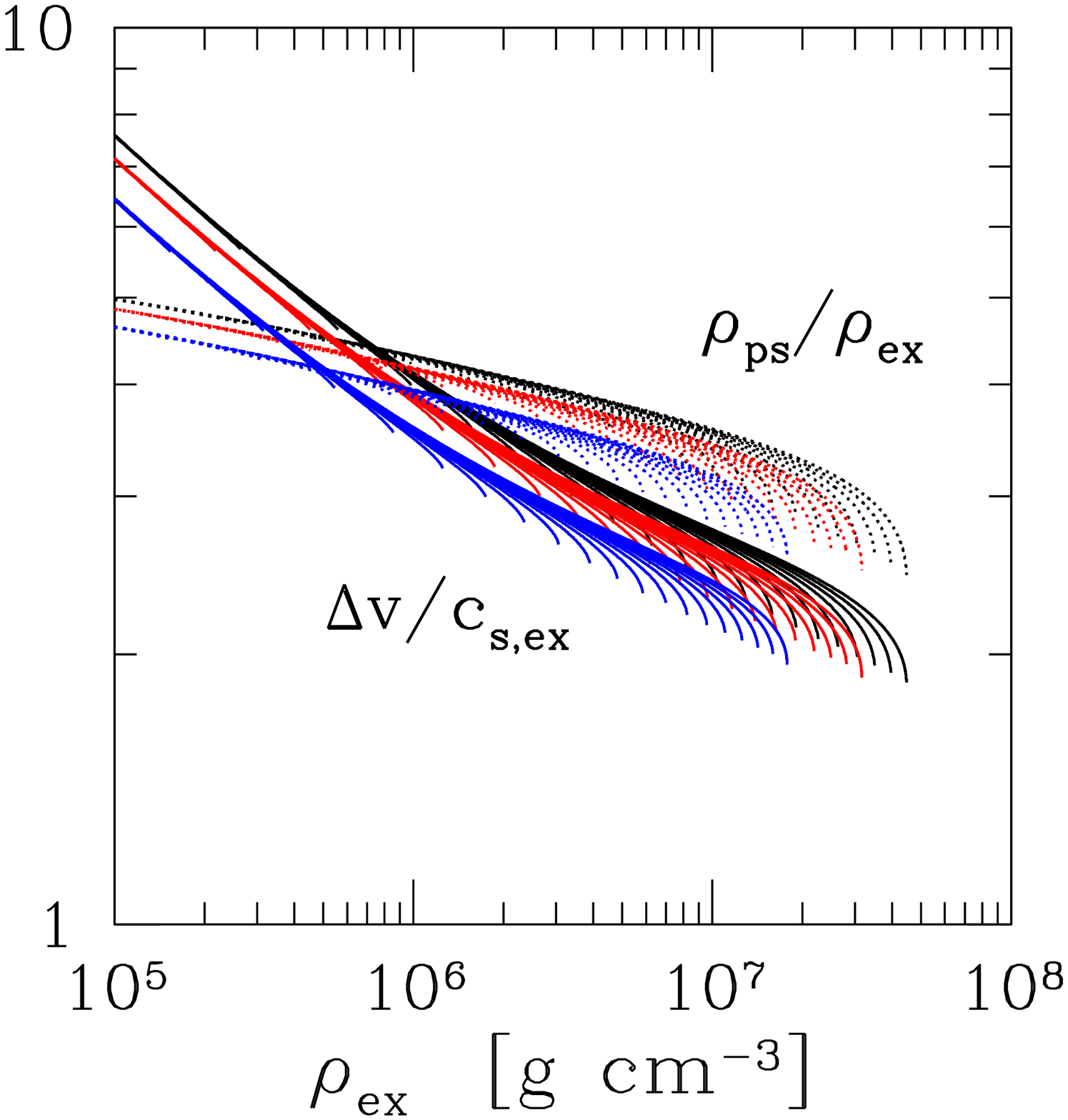}
\plotone{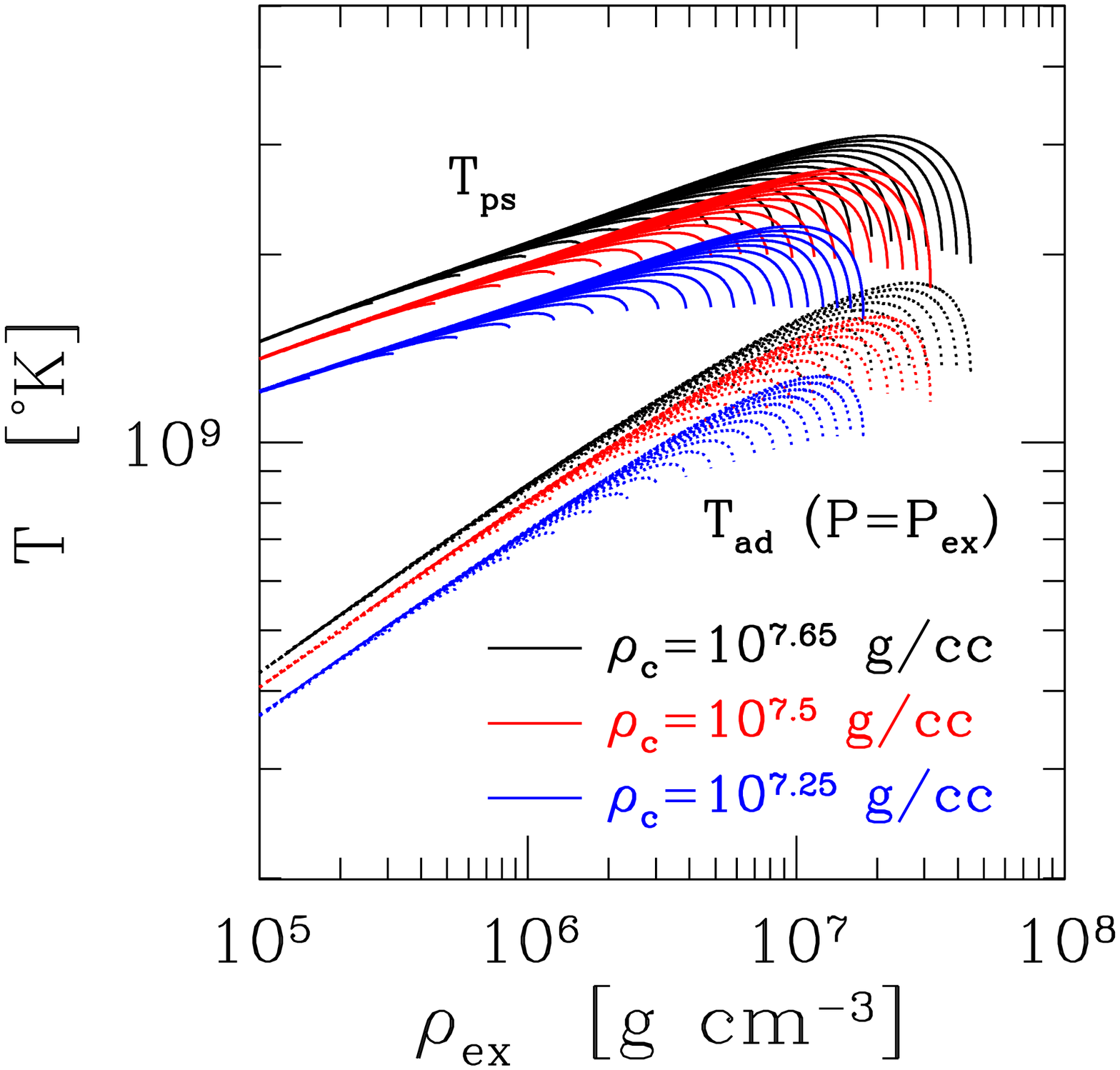}
\vskip .1in
\caption{{\it Top panel:}  Mach number and shock compression versus ambient stellar density for a LSD of mass 
${\cal M} = 4\times 10^{20}$ g and tension ${\cal T} = 10^{-8}c^4/G$.
Each color shows a sequence of LSD trajectories with a range of impact parameters and a single C/O WD mass
(black:  central density $\rho_c = 10^{7.65}$ g cm$^{-3}$ and mass $M_{\rm wd} = 1.07\,M_\odot$;  red:  $\rho_c = 10^{7.5}$
g cm$^{-3}$ and $M_{\rm wd} = 1.02\,M_\odot$;  blue: $\rho_c = 10^{7.25}$ g cm$^{-3}$ and $M_{\rm wd} = 0.92\,M_\odot$).
Squared impact parameter $(b/b_{\rm col})^2$ ranges linearly from $0$ to $0.95$ in steps of $0.05$.  Each trajectory
is followed until the first radial turning point inside the WD, where $v_r$ changes from negative to 
positive.  Beyond this point, the LSD is slowed sufficiently that burning is inefficient.
The top curve in each sequences corresponds to a zero angular momentum trajectory which reaches the 
center of the star on the first pass.  {\it Bottom panel, solid curves:}  post-shock temperature; {\it dotted curves:} 
temperature following adiabatic decompression to the ambient stellar pressure.}
\label{fig:mach}
\end{figure}

\subsection{Carbon  Burning Behind the Shock}

To calculate the post-shock flow, we iteratively solve the equations of mass, momentum and energy conservation
across the shock using the Helmholtz equation of state $P(\rho,T)$ based on a one-component plasma description of the ions
\citep{ts00}
\ba
P_{\rm ps} + \rho_{\rm ps}  v_{\rm ps}^2 &=& P_{\rm ex} + \rho_{\rm ex}\Delta v^2; \nn
{1\over \rho_{\rm ps}}\left(e_{\rm ps} + P_{\rm ps}\right) + {1\over 2}v_{\rm ps}^2 &=& 
{1\over\rho_{\rm ex}}\left(e_{\rm ex} + P_{\rm ex}\right) + {1\over 2}\Delta v^2.\nn
\ea
Here the subscript `ps' denotes post-shock and $e_{\rm ps}$, $v_{\rm ps} = (\rho_{\rm ex}/\rho_{\rm ps})\Delta v$ 
are total internal energy per unit volume and velocity in the frame of the shock.

The mach number and compression at the head of the shock are plotted in Figure \ref{fig:mach}
(top panel) for a range of impact parameters, and for three WD models with central densities
$10^{7.25}$, $10^{7.5}$, and $10^{7.65}$ g cm$^{-3}$ (masses $0.92$, $1.02$ and $1.07\,M_\odot$).  
The shock is strong near the entry point (low stellar mass density), and begins to weaken
near the first radial turning point, which coincides with the highest-density extension of the plotted curves.
The shock remains relatively weak during the inspiral that follows, so that peak burning rates are reached
before the first turning point.  

In WD models expected to contain a significant abundance of carbon ($M_{\rm wd} \lesssim 1.07\,M_\odot$; \citealt{farmer15}), 
the peak post-shock temperature 
(bottom panel of Figure \ref{fig:mach}) is high enough to trigger strong carbon burning.
However, the duration of peak compression is very short, only $\sim R_{\rm mag} / v_{\rm ps}$. 
We must therefore compare this hydrodynamic timescale, over which the shocked fluid decompresses back
to the ambient stellar pressure, with the ignition time $t_{\rm ign,c+c}$ for carbon burning.  The 
ignition time is defined here as the time to fully deplete the carbon, and we use the fitting formula
for isobaric burning provided by \cite{dursi06}.  In practice the electrons in the shocked white dwarf material 
are only moderately degenerate, and one finds $t_{\rm ign,c+c} \sim 0.3 e_{\rm th,ps}/\rho_{\rm ps}\dot\varepsilon_{c+c}$,
where $\dot\varepsilon_{c+c}$ the rate of energy release by carbon burning per unit mass, and $e_{\rm th}$ 
is the thermal energy density.

\begin{figure}
\epsscale{1.1}
\plotone{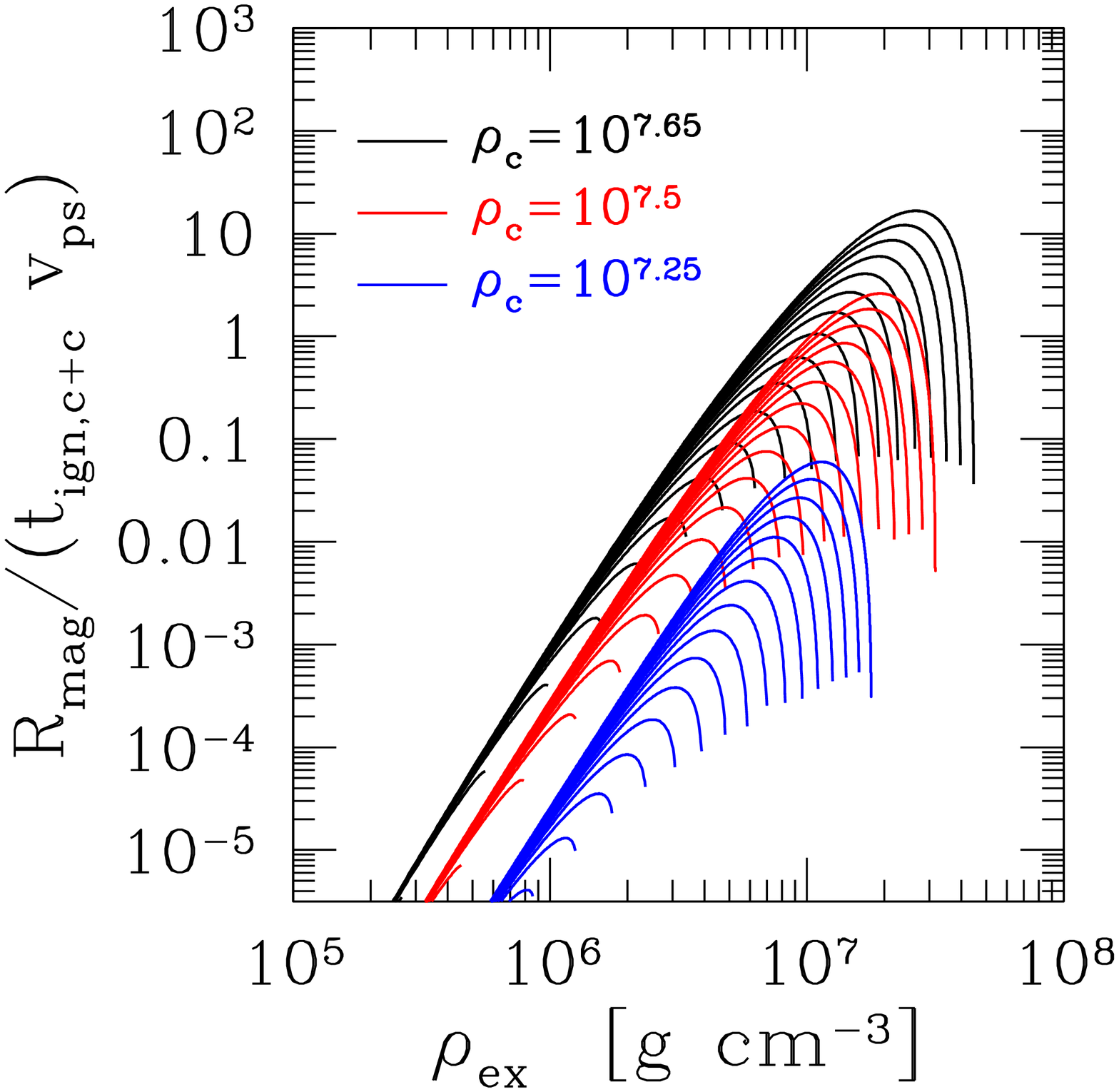}
\plotone{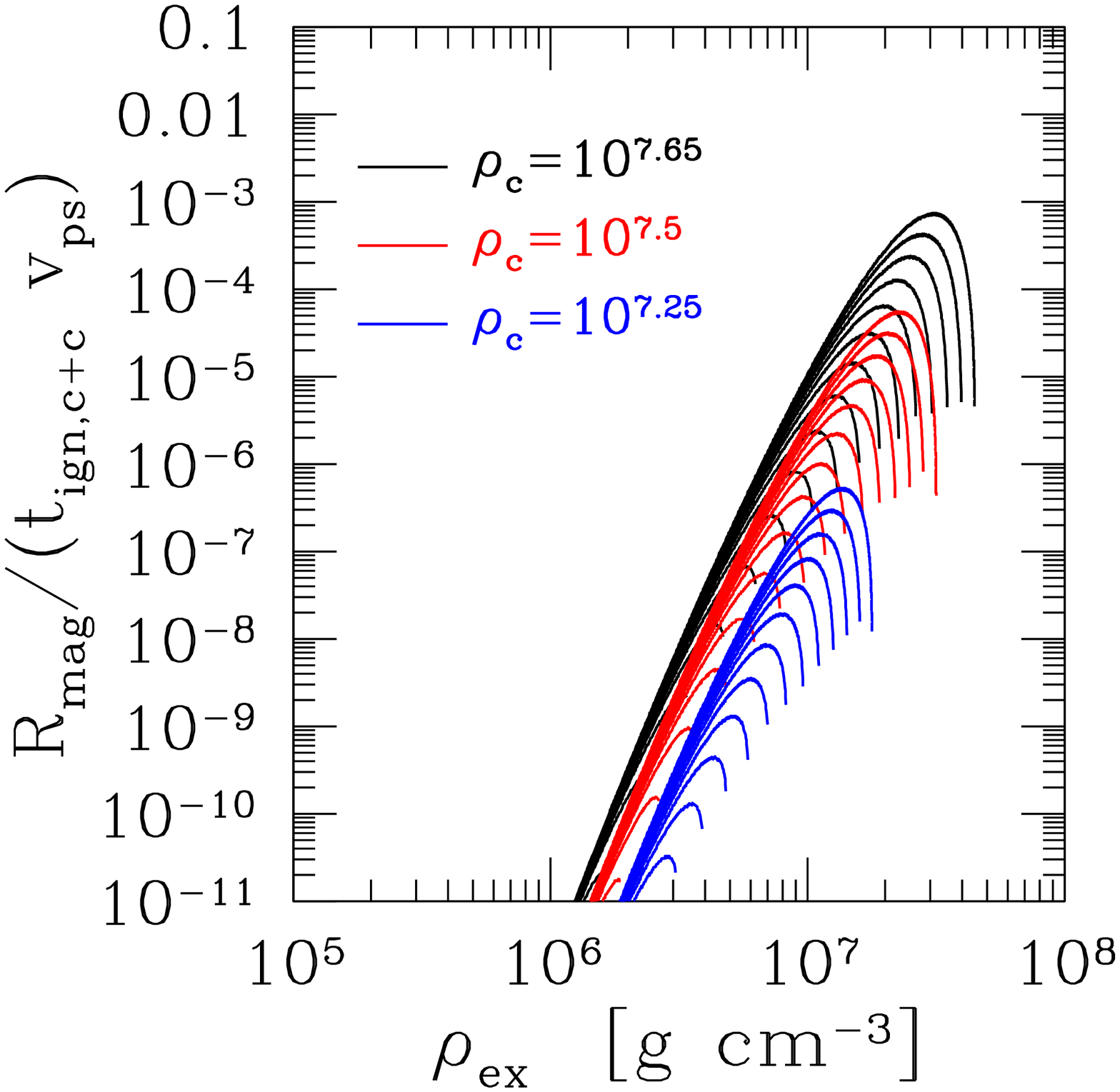}
\vskip .1in
\caption{{\it Top panel:}  Post-shock flow time $R_{\rm mag}/V_{\rm ps}$ compared with the ignition timescale 
$t_{\rm ign,cc}$ for carbon burning.  WD is assumed to have equal mass fractions of carbon and oxygen.  
Colors and lines correspond to the same LSD trajectories and WD models as plotted in Figure \ref{fig:mach}.
{\it Bottom panel:}  Same as top panel, but now showing how adiabatic cooling from the post-shock condition 
down to the ambient stellar pressure decreases the rate of carbon burning.  Each LSD shock trajectory is shown to a maximum
density (minimum radius) corresponding to the first turning point.}
\vskip .2in
\label{fig:tcc}
\end{figure}

The top panel of Figure \ref{fig:tcc} shows that carbon burning significantly raises the internal heat of the shocked
fluid in the heaviest ($1.07\,M_\odot$) WD model, but not in the lightest model ($0.92\,M_\odot$).   
In the first case, one can expect turbulent mixing of the heated carbon ashes with surrounding
unburnt stellar material to trigger sustained burning.  The  $1.02\,M_\odot$ model just passes the threshold
for sustained burning.

The next step is to follow the shocked fluid as it decompresses back to the ambient stellar pressure
(bottom panel of Figure \ref{fig:tcc}).  
First consider the case of slow post-shock burning.  Approximating the decompression as adiabatic,
one obtains the bottom set of temperature curves in Figure \ref{fig:mach}.
In the $0.92\,M_\odot$ model, one finds that the energy input from burning drops
below the loss from neutrino cooling.

It is also interesting to 
ask what happens when there is enough temperature growth to keep the heating rate within
the decompressed fluid faster than neutrino cooling.   Here it is important to keep in mind that the motion of the
LSD through the star creates strong velocity shear transverse to this motion.   Hydrodynamic turbulence will be
excited by this shear, and provide greatly enhanced transport of momentum and heat away from the center-of-mass 
trajectory.  This mixing process must slow down as the turbulent `trail' expands, but it is limited by the stellar stratification only after it has reached the relatively large scale $\sim (r R_{\rm mag})^{1/2}$.  This implies
a strong dilution of the heat deposited by the shock, and it is plausible that self-sustained thermonuclear burning 
will not be attained unless strong burning occurs in the immediate post-shock flow.  The range of C/O WD masses that
experience sustained burning would be broadened slightly if this conclusion were incorrect.

\begin{figure}
\epsscale{1.1}
\plotone{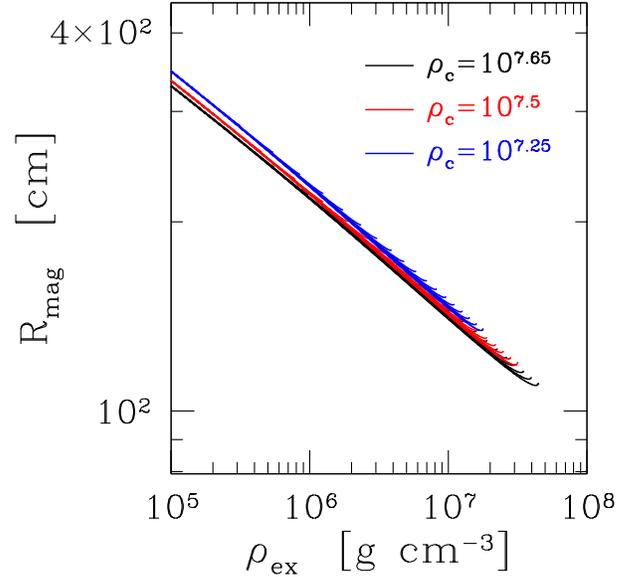}
\vskip .1in
\caption{Magnetospheric radius $R_{\rm mag}$ (Equation (\ref{eq:rmag})) of a LSD infalling through
WDs of various masses.   LSD and WD parameters are the same as in Figure \ref{fig:mach} and \ref{fig:tcc}.}
\vskip .2in
\label{fig:rmag}
\end{figure}

\begin{figure}
\epsscale{1.1}
\plotone{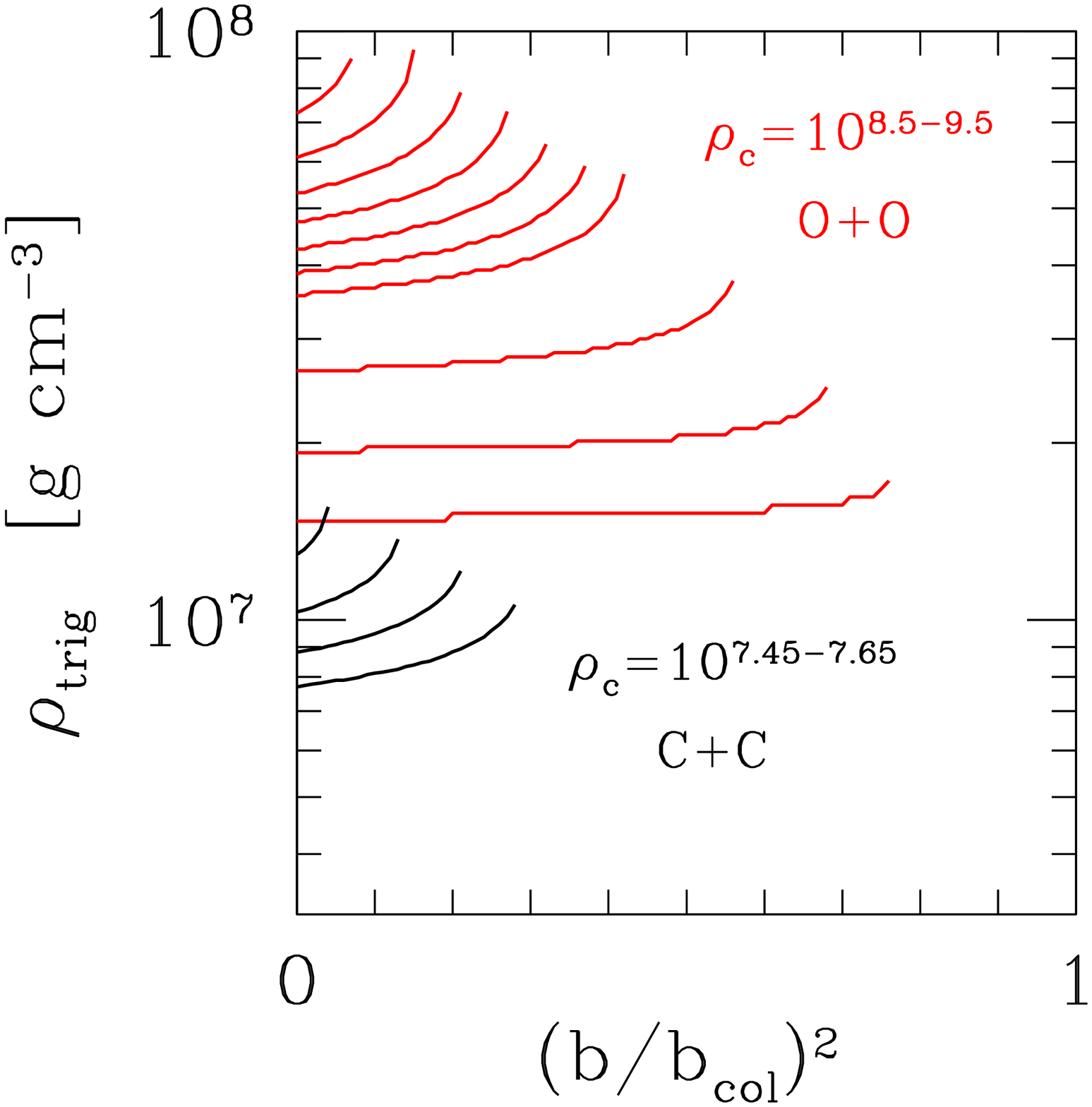}
\plotone{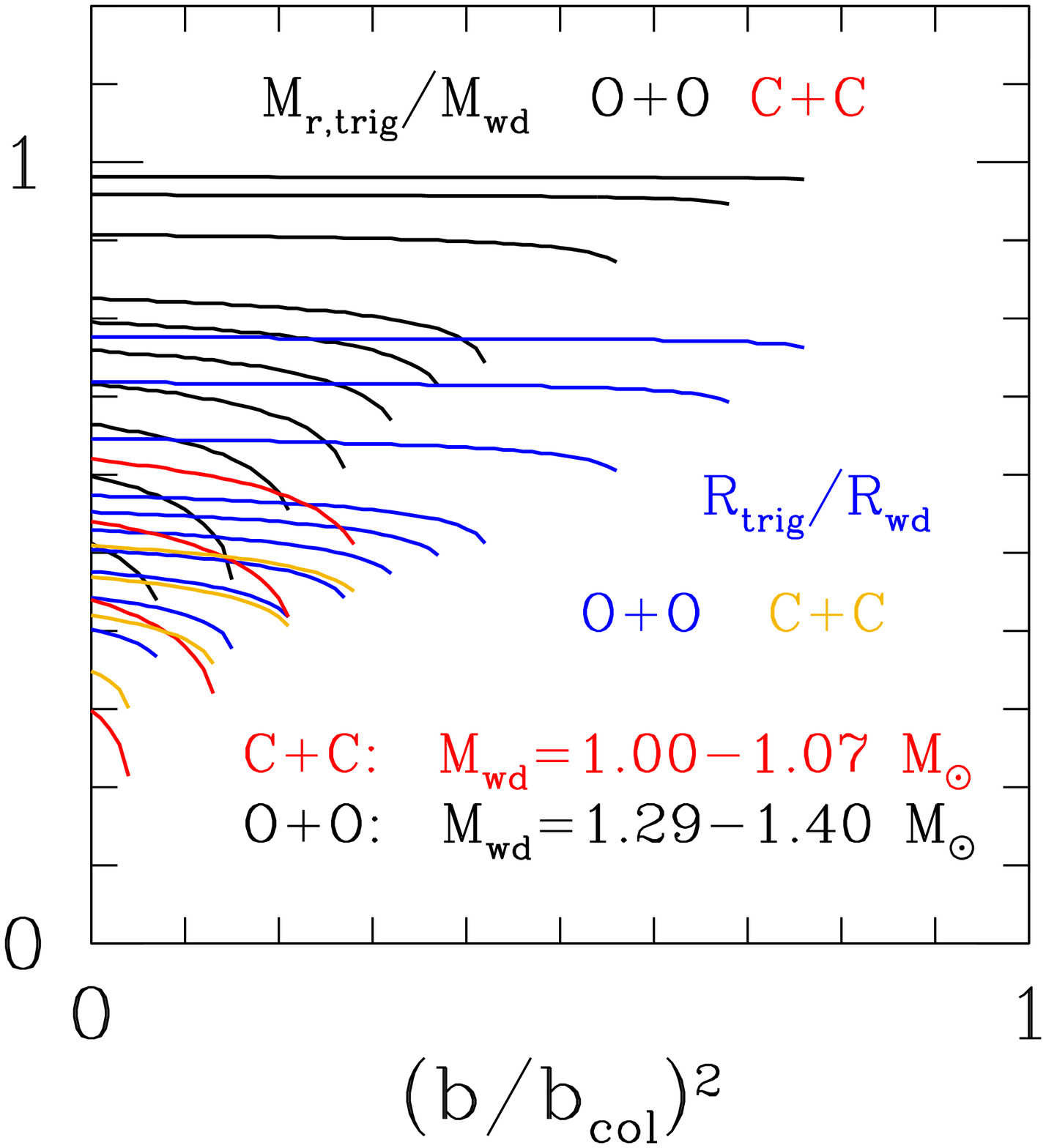}
\vskip .1in
\caption{{\it Top panel:}  Density at which sustained thermonuclear burning is first triggered along the trajectory of
LSD falling into a WD.  For a given composition, this density is {\it lower} for WD models of a higher mass
and compactness, corresponding to a relatively large radius and enclosed mass.  {\it Black lines:}  cold C/O 
WD models of central density $10^{7.45-7.65}$ g cm$^{-3}$, increasing in steps of 0.05 in $\log_{10}\rho_c$.  
Corresponding WD masses range from 1.0 to 1.07$\,M_\odot$.  {\it Red lines:} ONeMg WDs of central density 
$10^{8.5-9.5}$ g cm$^{-3}$, increasing in steps of 0.05 and then 0.25 in $\log_{10}\rho_c$.  Corresponding WD
mass ranges from 1.29 to 1.40$\,M_\odot$.  {\it Bottom panel:}  Outer radius $R_{\rm trig}$ and enclosed mass 
$M_{r,\rm trig}$ of sustained burning.}
\vskip .2in
\label{fig:trigger}
\end{figure}

\subsection{Oxygen Burning in Massive ONeMg WDs}

The calculations in the preceding Section were repeated for model WDs of ONeMg composition.   The
higher threshold temperature for runaway oxygen burning implies a higher mass threshold for
the condition $t_{\rm ign,oo} = R_{\rm mag}/V_{\rm ps}$.   We find that $M_{\rm wd} > 1.29\,M_\odot$
is required, with a weak dependence on the velocity-at-infinity of the infalling LSD.

\subsection{Off-Center Spark}

The long history of modeling thermonuclear explosions of WDs suggests that the 
observational consequences of a deflagration depend not only on the stellar mass, but also
on the position of the initial spark.   Figure \ref{fig:trigger} shows that this spark is
situated at moderate depths when a LSD falls into a C/O WD of mass $1.0-1.07\,M_\odot$,
but at relatively shallow depths for more mass ONeMg WDs heavier than $1.29\,M_\odot$.  In 
fact, if the composition of the WD is fixed, then self-sustained burning is first triggered 
at lower {\it absolute} densities as the stellar mass is raised.  
That is because the greater compactness of the more massive WD 
implies a stronger drag at a given density.  

\subsection{Prompt Detonation vs. Deflagration}\label{s:prompt}

Does the shocking of WD material by an infalling LSD produce a deflagration (self-sustained burning by
a turbulent flame) or a direct detonation (complete burning behind a propagating shock)?   Complete
burning within the hydrodynamic disturbance produced by the LSD (e.g. the top black curve of Figure \ref{fig:tcc}) 
is not a guarantee that a detonation will be triggered.  That is because 
a self-sustained detonation is suppressed by shock curvature \citep{niemeyer97,dursi06}.
To evaluate the role of curvature here, we compare the magnetospheric radius $R_{\rm mag}$, shown
in Figure \ref{fig:rmag} for the same trajectories and WD models as Figures \ref{fig:mach} and \ref{fig:tcc},
with the minimum hotspot size calculated by \cite{dursi06} (see their Equation 11):
\be\label{eq:rdet}
R_{\rm det} = 1.63\times 10^3\,\rho_{\rm ex,8}^{-1.22}\quad{\rm g~cm^{-3}}.
\ee

One sees that $R_{\rm mag}$ falls short of expression (\ref{eq:rdet}) by some two orders of magnitude at 
the density $\sim 10^7$ g cm$^{-3}$  where burning is triggered in our C/O WD models (Figure \ref{fig:trigger}).    
A related effect is that the shock speed (Figure \ref{fig:mach}) rises above the 
Chapman-Jouget speed (the approximate propagation speed of a self-sustained detonation) 
at the point of strongest burning.  As a result, the standoff of the shock from the magnetospheric
boundary of the LSD cannot be strongly inflated by the heat input from burning.   

We conclude that the disturbance produced by an infalling LSD results in localized production of
iron group elements in $1.0-1.07\,M_\odot$ WDs, but the case for a prompt detonation is uncertain.
The spark has a very different (approximately cylindrical) geometry to that usually assumed.

This (admittedly exotic) process gives a concrete example of an interesting problem which deserves
further exploration through simulation:  impulsive and directional forcing of thermonuclear burning, as opposed
to triggering by an initially static hotspot (the more commonly considered problem).

We discuss the implications of this triggering model for iron group nucleosynthesis, the Type Ia supernova rate,
and the nature of the Type Ia progenitors, in the concluding Section \ref{s:wdc}.

\section{Conclusions and Predictions}\label{s:summary}

We have described the collisional properties of primeval magnetic dipoles of $\sim 1$ mm
size and mass $\sim 10^{20}$ g in an evolving cold dark matter dominated universe.   
The cross section for free classical dipoles to collide has been calculated.   If such
LSDs comprise a significant component of the dark matter then their collisions produce intense, 
low-frequency electromagnetic pulses at a rate consistent with the observed rate of FRBs.  
Superconducting `spring' loops with tension ${\cal T} \sim 10^{-8}c^4/G$ are an
interesting microphysical candidate for LSDs.  Young SMBHs forming by direct collapse of gas
in dark matter halos of medium mass will be surrounded by long-lived collisional rings of LSD.
These objects are repeating sources of FRBs, and may dominate the global rate.  The mass distribution
of LSDs is significantly modified by repeated collisions in such an environment.  
We apply our results to the repeating and recently localized source FRB 121102.

There are a number of additional observational constraints on LSDs, and interesting consequences of
the web of physical processes examined here.  These include a promising
application to the triggering of thermonuclear explosions of WDs colliding with LSDs:  the same LSD properties
that explain the energies and rates of FRBs also imply an explosion rate of $> 1\,M_\odot$ WDs, and 
a corresponding yield of $^{56}$Fe, that are remarkably close to those of Type Ia supernovae.  

In addition, the cosmic abundance of LSDs is squeezed from above 
the mass range preferred for FRB emission by microlensing constraints, and from below
by a high rate of disruptive collisions with WD stars.  We make testable predictions of
bright and narrow bursts at high ($\sim 100$ GHz) frequencies associated with nearby supermassive
black holes, and note that $\mu$Jy level radio emission with enormous proper motion is potentially
detectable from the closest LSDs moving at high speed through the outer Solar system.  
The approach to the FRB puzzle advanced here is testable from multiple directions.

\subsection{Constraints from Host Dispersion Measure, Rotation Measure, and Pulse Scattering 
            Broadening}\label{s:dm}

An intense GHz electromagnetic pulse with energy $\sim 10^{39}$ erg 
can avoid synchrotron absorption within the inner accretion flow of a SMBH (mass $M_\bullet$) 
by raising the energies of the flow electrons.  Plasma dispersion in such a dense medium
spreads out the pulse in the propagation direction, thereby reducing the relativistic quiver radius to 
\ba\label{eq:rquiv}
R_{\rm rel,\omega} &\sim& \left({\omega{\cal E}_\omega\over 4\pi n_e m_ec^2}\right)^{1/3} \nn
                    &=& 4.6\times 10^{12}\,(\omega{\cal E}_\omega)_{39}^{1/3} n_{e,6}^{-1/3}\quad{\rm cm}
\ea
(Equation (120) of Paper I).  In our approach to the emission mechanism, the effect is dominated
by the peak of the spectrum at $\sim 10^2$ GHz, where one may expect $\omega {\cal E}_\omega$ to 
reach $10^{40-41}$ erg.  

The wave grows too weak to induce relativistic motion at the radius (\ref{eq:rquiv}).  Synchroton 
absorption of GHz photons can be avoided in the relatively undisturbed accretion flow just
outside this point if
\be\label{eq:nemax}
n_{e,\rm ISCO} < 2\times 10^7\,\nu_9^{8/9}(\omega{\cal E}_\omega)_{41}^{4/9} M_{\bullet,6}^{-14/9}
\quad {\rm cm^{-3}}
\ee
(Equation (132) of Paper I).  This estimate is based on recent models of radiatively inefficient accretion flows 
(RIAFs) which find a profile $n_e(r) \propto r^{-1}$ in between the Bondi radius $R_B$ and the ISCO,
with the heated electrons become non-relativistic at a distance $R_{\rm rel} \sim 200R_g$ from the hole
\citep{yuan03,yuan12}.   

\subsubsection{Dispersion}

The accretion flow will therefore contribute negligibly to the dispersion measure (DM) of 
observable FRBs emitted by LSD collisions near the ISCO of a SMBH (compare with
the observed ${\rm DM} \sim 300$-$500$ cm$^{-3}$ pc):
\ba
{\rm DM} &\sim& n_{e,\rm ISCO} R_{\rm ISCO} 
\ln\left({R_B\over R_{\rm rel}}\right)\nn
 &\sim& 10\, n_{e,\rm ISCO,7} M_{\bullet,6}\quad{\rm cm^{-3}~pc}.
\ea
Here $R_B \sim 1\times 10^{17}M_{\bullet,6}$ cm given a plasma temperature $\sim 1$ keV in the
circumnuclear region.  

\subsubsection{Faraday Rotation}

Strong linear polarization is a natural consequence of the three emission channels for a superluminal
electromagnetic wave that are described in Paper I.   The starting point is a relativistic, magnetized shell
with a dilute gas of embedded warm $e^\pm$ pairs that was created during a collision between LSDs.
A superluminal wave is emitted by
\vskip .05in\noindent
1. Direct linear conversion of the magnetic field embedded in the shell;
\vskip .05in
\noindent
2. Reflection of an ambient magnetic field by the surface of the conducting shell (see also 
\cite{blandford77} for a discussion of radio emission from exploding black holes);
\vskip .05in
\noindent
3. Excitation of an electromagnetic wave by the corrugated shell surface, with the corrugations 
being triggered by reconnection of the shell magnetic field with an ambient field.
\vskip .05in

A relatively large rotation of this linear polarization would be produced by propagation through the
surrounding RIAF.   The contribution to the rotation measure (RM) from the RIAF is concentrated at 
radius $\sim R_{\rm rel}$.  Given that the magnetic pressure scales as a constant fraction of the
thermal electron pressure ($\sim r^{-2}$), one gets
\be\label{eq:RM}
{\rm RM} \sim 10^6 \; n_{e,\rm ISCO,7} M_{\bullet,6}\left({R_{\rm rel}\over 10^3\,R_g}\right)^{-1/2}
\quad {\rm m^{-2}}.
\ee
This appears to be near the threshold of detectability of existing FRB measurements;  a much larger RM would
presumably cause depolarization at $\sim$ GHz frequencies.

An important test of these ideas comes from FRBs which show low RMs, the main example so far being FRB 150807 
\citep{ravi16}.  Our approach requires that these events arise from LSD collisions in galactic
halos, far removed from SMBHs.  Unless a mechanism can be found for suppressing the accretion rate
onto a SMBH well below the already low level of Sgr A*, there is a strong requirement that a FRB showing
low RM will not repeat.

By the same token, the absence so far of detected polarization in the repeating bursts of FRB 121102
requires Faraday depolarization.

\subsubsection{Scattering Delay}

Although a high level of turbulence is expected within a RIAF, the scattering delay of an electromagnetic
pulse emitted in its inner parts would be suppressed by the short path length:
\be
\delta t \sim {1\over 2}\left({\omega_{Pe}\over\omega}\right)^4{r\over c}
         \sim 0.3\;n_{e,\rm ISCO,7}^2 \nu_9^{-4} M_{\bullet,6}\quad{\mu\rm s}.
\ee
The dominant contribution comes once again from radius $\sim R_{\rm rel}$.  Here we have normalized 
$n_{e,\rm ISCO}$ to the limiting value (\ref{eq:nemax}) allowed by synchrotron absorption.

\subsection{Repetitions -- Application to FRB 121102}

We now describe how the results reported in this paper and Paper I may be applied to the repeating FRB
121102.  This source has been localized to a low-mass dwarf galaxy at redshift 0.19 \citep{tendulkar17}, 
and appears to sit within $\sim 40$ pc of a compact radio source \citep{marcote17,chatterjee17}.

Obtaining FRB repetitions by the
radiation mechanism advanced here must involve a very compact cloud of LSD.  This we suggest forms most
easily by direct collapse of gas combined with a small amount of dark matter into a SMBH (Section \ref{s:smbh}).  
It is natural to identify such a SMBH with the radio counterpart of 121102.  
The low stellar mass ($\lesssim 10^8\,M_\odot$) of the host galaxy of FRB 121102 suggests that this SMBH is
of modest mass, $\lesssim 10^6\,M_\odot$.   Equations (\ref{eq:nd}) and (\ref{eq:ndeq}) then show that the 
annihilation rate of LSDs has relaxed to the equilibrium inflow rate as determined by gas drag.   

We require a low present accretion rate onto the black hole for the high-amplitude radio pulses to escape
unattenuated.  Indeed, a SMBH mass below $10^6\,M_\odot$ could significantly improve the transparency of
the accretion flow, as described in Section \ref{s:dm}.   Although the radio source around
the SMBH could not be powered by the present accretion, one recalls that strongly intermittent accretion
onto Sgr A$^*$ has been inferred from measurements of X-ray reflection, with a major outburst occuring within
the last $\sim 10^2$ yr \citep{churazov17}.  The radio source associated with FRB 121102 could represent
persistent synchrotron emission from a similarly recent outburst.

We suppose that the SMBH has been accreting dark matter from an adiabatically contracted cusp for 
a time $t_{\rm SMBH}$.  The density profile of this cusp is given by Equation (\ref{eq:rhod0}).  We set
the outer radius $r_{d,\rm acc}$ of the accreted dark matter particles by equating $t_{\rm SMBH}$ with
the drag time (\ref{eq:tdrag3}), finding
\be
{r_{d,\rm acc}\over R_{\rm ISCO}} = 1.8\times 10^4\,\left({t_{\rm SMBH}\over 3~{\rm Gyr}}\right)^{6/7}
{f_{\rm em}^{2/7} {\cal R}_{-1}^{6/7} \dot m_{-3}^{4/7}\over f_{\rm hydro}^{6/7} {\cal M}_{20}^{4/7}
 M_{\bullet,6}^{4/7}}.
\ee
Here $\dot m$ is the long-term average of the accretion rate onto the SMBH, as measured in terms of the
Eddington rate.  The long-term average of the inflow rate of LSDs to the ISCO is, then,
\ba\label{eq:supply}
\dot N_{d+} &=& {1\over t_{\rm SMBH} {\cal M}} {dM_d\over d\ln r_d}\biggr|_{r_{d,\rm acc}}\nn
            &=& 5\,\left({t_{\rm SMBH}\over 3~{\rm Gyr}}\right)^{2/7}
  {M_{\bullet,6}^{15/7} \dot m_{-3}^{6/7} f_{\rm em}^{3/7}{\cal R}_{-1}^{9/7}\over 
       f_{\rm hydro}^{9/7}{\cal M}_{20}^{13/7}}\quad{\rm yr}^{-1}.\nn
\ea
Here we have normalized the seed mass profile during the early gas collapse phase
to $\sigma_d = 40$ km s$^{-1}$, $\sigma_g = 20$ km s$^{-1}$ and transition radius $R_{\rm gas} = 100$ pc
between gas and dark matter (Equation (\ref{eq:rhod0})).

A supply rate of $\sim 10$ yr$^{-1}$ of LSD to a SMBH in the FRB 121102 host may be too small, given
the modest duty cycle of the radio monitoring, even allowing for repetitions produced by gravitational lensing
and reflection off dense plasma (Section \ref{s:rep}).   The LSD mass normalization made in this estimate
corresponds to a GHz burst energy $\sim 10^{39}$ erg, and could easily be reduced by an order of magnitude
in this environment.  We showed in Section \ref{s:smbh} that repeated collisions between LSDs orbiting the SMBH
create a high-energy tail to the burst energy distribution.  Substituting ${\cal M} \sim 10^{19}$ g in  
Equation (\ref{eq:supply}) then implies an increase to several hundred bursts per year.  There are, however,
compensating constraints associated with an increased collision rate with WD stars (Section \ref{s:wd3}).

\subsubsection{Repetition Timescales}\label{s:rep}

Repetitions from FRB 121102 do not appear to happen in a Poissonian manner.   A lower timescale for
repetition is obtained from the differential gravitational lensing delay, which is of the order
\be\label{eq:tlens}
\Delta t_{\rm lens} \sim {R_{\rm ISCO}\over c} = 30\,M_{\bullet,6}\quad{\rm s}.
\ee
This is comparable to the shortest repetition period observed from FRB 121102 (22.7 seconds; \citealt{spitler16})
for a $\sim 10^6\,M_\odot$ black hole.  

Longer repetition periods are made possible by a geometrical delay associated with reflection.  There are
several strong constraints on reflection as a source of bright and narrow FRB `afterimages', all of which
are plausibly satisfied by an accretion flow onto a SMBH.
\vskip .05in \noindent
1. The radius of curvature $R_c$ of the mirror must be comparable to its separation $D_{sm}$ from the FRB source.
Otherwise the brightness of the reflected pulse is greatly diluted, by a factor $\sim (R_c/D_{sm})^2$.  
\vskip .05in \noindent
2. The plasma density must reach 
\be\label{eq:neref}
n_{e,\rm ref} \sim 1\times 10^{10}\theta_{\rm ref}^2\nu_9^2\quad {\rm cm^{-3}}
\ee
in order to reflect GHz radiation through an angle $\theta_{\rm ref}$.
\vskip .05in \noindent
Neither of these constraints is readily satisfied by an intervening plasma cloud within the host galaxy.  
By contrast, a warm accretion disk orbiting a SMBH can easily reach such a density, and a warped disk
naturally has curvature radius $R_c \sim D_{sm}$ at a distance $D_{sm}$ from the hole.  Of course, the
reflection density (\ref{eq:neref}) greatly exceeds the maximum $n_e$ (\ref{eq:nemax}) that would allow the
pulse to escape from the ISCO.  This might point to `magnetic arrest' of the innermost accretion flow by a
concentrated magnetic flux bundle, to a two-phase structure of the flow, or to a radial transition from an outer thin
disk to an inner hot and slowly cooling RIAF (Section \ref{s:mirror}).

\vskip .05in \noindent
3. Reflection cannot occur too close to the FRB source:  electrons overlapping the pulse are heated to
$\langle\gamma_e\rangle \gtrsim m_p/4m_e$ inside the radius $R_{\rm rel,\omega} \sim 10^{13}$ cm (Equation
(\ref{eq:rquiv})).  A characteristic geometric delay is then $R_{\rm rel,\omega}/c \gtrsim 300$-1000 s.  This 
is, interestingly, comparable to the duration of two clusters of GHz pulses detected from FRB 121102 
\citep{spitler16,scholz16}.

Such a transition to a subrelativistic electron motion could begin only after the electromagnetic pulse
begins to overlap an intervening dense plasma screen.  Placing the screen at a distance $D_{sm} = 10^{13}$ cm 
from the FRB, the transition begins after the $\sim 0.01$-1 THz forward pulse penetrates to a column 
$\delta N_e \sim \omega{\cal E}_\omega/4\pi R_s^2m_ec^2 = 
1\times 10^{20} (\omega{\cal E}_\omega)_{41} D_{sm,13}^{-2}$ cm$^{-2}$ below the disk surface,
with non-relativistic electron energies being reached at a column $\ln(m_p/4m_e) \sim 6$ times greater 
(see Equation (127) of Paper I).

This approach to repetitions in FRB 121102 is testable with upcoming monitoring.  
Repetitions on a timescale much less than $\sim 10$ s, but much longer than a few ms, would invalidate
the hypothesis of gravitational lensing.  One should also observe a significant gap in the repetition
interval between the lensing time (\ref{eq:tlens}) and the minimum propagating time of a few
hundred seconds to a non-relativistic plasma mirror.

\begin{figure}
\epsscale{1.1}
\plotone{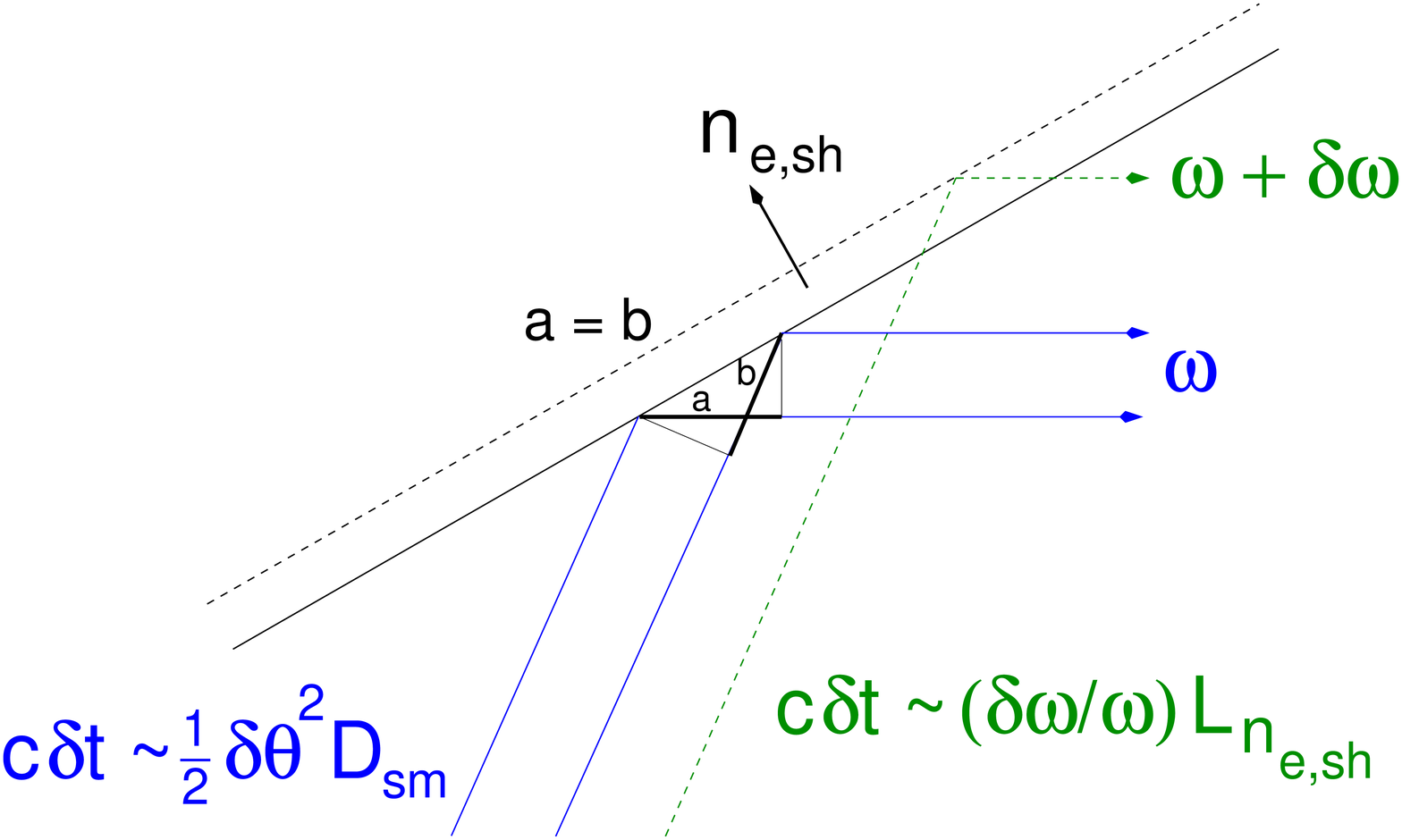}
\vskip .1in
\caption{Differential delay experienced by a collimated electromagnetic pulse reflecting off a plasma screen
at a distance $D_{sm}$ from the FRB, relative to the mean geometric delay $t_d = {1\over 2}\theta_{\rm ref}^2 D_{sm}/c$.  At a fixed
frequency (blue rays) the delay is dominated by the curvature of the wavefront.  The angle $\delta\theta$
has contributions from the finite emission size ($\delta\theta \sim \Gamma_{\rm em}^{-1} m$, where
$\Gamma_{\rm em}$ is the Lorentz factor of the magnetized shell at the emission point, and $m$ is the magnification
of the source by the mirror) and from corrugations of the mirror surface.  The differential delay between
neighboring frequencies is dominated by the electron density gradient within the mirror.  Here it is essential
to take into account the energization of electrons in the mirror by a narrow 0.01-1 THz precursor
of the GHz pulse.  A rapid transition from relativistic to non-relativistic electron energies occurs
at an electron column $\sim 10^{20}$ cm$^{-2}$ for a mirror-FRB displacement $D_{sm} \sim 10^{13}$ cm.
Here $n_{e,\rm sh}$ is the electron density in the plasma screen, and $L_{n_e}$ its gradient scale normal to the screen.}
\vskip .2in
\label{fig:delay}
\end{figure}

\subsection{Pulse Broadening During Reflection}\label{s:mirror}

In addition to the overall geometric delay, reflection creates a differential
delay between neighboring frequencies.  If the electrons in the screen remained subrelativistic, this would
potentially add to the dispersion measure in excess of the $\sim \delta{\rm DM}
\sim 3$-10 cm$^{-3}$ pc limit imposed by pulse-to-pulse measurement variations in FRB 121102.  There would
also be a differential delay (Figure \ref{fig:delay}) associated with the greater penetration of 
higher frequencies into the mirror.  For example, in an isothermal atmosphere with scaleheight $h$, the
geometric delay between frequencies $\omega$, $\omega + \delta\omega$ is
\be
\delta t_d = 4{h\over c}\ln\left(1+{\delta\omega\over\omega}\right)
\ee
in the case of normal incidence.
This only approximates a (negative) plasma dispersion over a narrow band.   Making that correspondence
between $h$ and $\delta{\rm DM}$, and demanding that the geometric delay not cause a measurable deviation from
the cold plasma dispersion law at finite bandwidth ($\delta\nu/\nu \sim 0.4$ in the Arecibo observations
of FRB 121102: \citealt{scholz16}), one would require the scale height to be smaller than
\be\label{eq:hmax}
h < {e^2\over 4\pi m_e \nu^2}\delta{\rm DM} = 
6\times 10^8\;\nu_9^{-2}\left({\delta{\rm DM}\over 10~{\rm cm^{-3}~pc}}\right)\quad{\rm cm}.
\ee

The scalelength of the plasma frequency variation is slightly shortened by the transition of
the electrons from relativistic to sub-relativistic motion within the strong electromagnetic pulse,
by a factor $\sim 1/\ln(m_p/4m_e)$.  Combining this effect with Equation (\ref{eq:hmax}), 
a rough upper bound on $h$ is $\sim 10^9$ cm for a 2 GHz pulse.  

Such a small scaleheight at the relatively low column $N_e \lesssim 10^{20}$ cm$^{-2}$ is clearly inconsistent
with a magnetically active accretion disk, in which the magnetorotational instability would generate a warm
corona with a larger column.  However, an accretion disk feeding a SMBH has a relatively low effective temperature,
so that mid-plane temperatures below $\sim 10^4 \,\K$ may be reached at $\sim 10^2-10^3$ gravitational radii
and low accretion rates.   Such a weakly ionized disk could hibernate in a state reminiscent of the quiescent
states of cataclysmic variables \citep{nayakshin03}.  Its presence would nicely complement the low electron
density (\ref{eq:nemax}) that we require within a hot RIAF in the FRB emission zone near the ISCO.

For example, the mid-plane of a thin gas pressure-dominated disk with opacity provided mainly by free-free absorption 
\citep{nt73} would reach the ionization threshold at an accretion rate
\be
{\dot M\over \dot M_{\rm edd}} = 0.005\,\alpha_{-2}^{2/3}\varepsilon_{\rm rad,-1}M_{\bullet,6}^{2/3}
\left({r\over 10^4\,R_g}\right)^{5/2}.
\ee
A temporary shut off in torque at $\sim 10^4$ gravitational radii due to weak ionization would allow the inner
parts of the disk to drain and also reach the ionization threshold, all the while remaining gravitationally stable.   

As the disk at radius $D_{sm} \sim 10^2\,R_g$ drops to a mid-plane temperature of $\sim 10^4\,\K$, its
mid-plane scaleheight reaches
\be
{h\over D_{sm}} \sim 4\times 10^{-4}\,T_4^{1/2}\left({D_{sm}\over 100\,R_g}\right)^{1/2}.
\ee
The vertically integrated mass column would drop to about $10^4\,M_{\bullet,6}^{4/5}$ g cm$^{-2}$ 
for the normalizations of viscosity parameter $\alpha$ and inner radiation efficiency $\varepsilon_{\rm rad}$ given
above.  A thin, externally ionized surface layer of electron column $\sim 10^{20}$ cm$^{-2}$ would sit at a
height $z_{\rm ion} \sim 6\,h$ above the mid-plane, and have a scaleheight
\be
{h_{\rm ion}\over D_{sm}} \sim  {h^2\over z_{\rm ion} D_{sm}} \sim 6\times 10^{-5}\;T_4^{1/2}
                                 \left({D_{sm}\over 100\,R_g}\right)^{1/2}.
\ee
The hydrostatic pressure in such a cool surface layer at the reflection density (\ref{eq:neref})
is comparable to the external pressure of a RIAF with density $n_{e,\rm ISCO} 
\lesssim 10^7$ cm$^{-3}$ at the ISCO and scaling as $n_e\kB T_e \propto r^{-2}$ at larger radius.  
It is also straightforward to check that heating by Coulomb scattering warm electrons that
penetrate the cool surface layer from the RIAF is not able to overcome in situ atomic cooling 
at a column $\gtrsim 10^{18}$ g cm$^{-2}$. 

This quiescent surface scale height is marginally consistent with (\ref{eq:hmax}), suggesting that the duration
of the reflected pulse will be determined by the geometric delay during reflection.
In fact, the electron density in the ionized surface layer would reach the threshold for reflecting
GHz waves at a column somewhat lower than the column that would be heated to relativistic energies
by the highest-frequency part of the electromagnetic pulse.  As a result, reflection of the GHz wave 
would occur at a depth in the mirror where the electrons were still mildly relativistic.  The reflected wave
frequency would be slightly modified by the bulk motion induced in the mirror electrons.  

Additional sources of geometric delay come from the finite size of the emitting patch on the relativistic
shell produced by the tiny FRB explosion.  Consider the simplest case
where the mirror does not magnify the electromagnetic pulse. Then the solid angle of the observable patch
of the fireball surface is $\sim \pi/\Gamma_{\rm em}^2$,
where $\Gamma_{\rm em}$ is the fireball Lorentz factor at the emission radius of the GHz photons.  
One finds $\Gamma_{\rm em} \sim 10^4$ from Equations (53) and (110) of Paper I.
Two rays of the same frequency emitted from opposite sites of the observable patch experience a differential
delay between emission and reflection
\be\label{eq:fs}
\Delta t_d \sim {D_{sm}\over \Gamma_{\rm em}^2 c} \sim 3\times 10^{-6}\;\Gamma_{\rm em,4}^{-2} D_{sm,13}\quad {\rm s}.
\ee
This would become observationally significant only if the plasma mirror magnified the GHz image,
in which case $\Delta t_d$ in Equation (\ref{eq:fs}) would be increased by a factor $m^2$, where $m$ is
the magnification.  Broadening could also arise from a high-wavenumber corrugation of the screen surface,
through an angle $\delta\theta > \Gamma_{\rm em}^{-1}$.

\subsection{Cosmic Redshift Distribution of \\ Electromagnetic Pulses}

Rare collisions between LSD particles in galactic halos are distributed broadly in cosmic redshift:
we find a fairly uniform rate of $4\times 10^4$ yr$^{-1}$ per comoving Gpc$^3$ between $z = 0$ and 2 for the
LSD parameters of Figure \ref{fig:rate} and \ref{fig:rate_halo} ($f_{\rm LSD} = 1$ and ${\cal T} = 10^{-8}c^4/G$).

On the other hand, collisions within LSD rings orbiting supermassive
black holes are concentrated closer to the epoch when these black holes form by direct collapse
of gas clouds (Section \ref{s:smbh}).   This source of LSD collisions should be
concentrated toward the redshift of peak SMBH assembly ($z \sim 2$ as measured by bright AGN activity).  
Unfortunately a high annihilation rate driven by a high accretion rate onto a SMBH
would also be accompanied by strong synchrotron absorption at GHz frequencies (Equation (\ref{eq:nemax})
and Paper I).

\subsection{Narrow and Ultra-Luminous 100 GHz Pulses from \\
Nearby Supermassive Black Holes}\label{s:pred}

The collision of two LSDs of mass ${\cal M} \sim 10^{20}$ g and size ${\cal R} \sim 0.1$ cm releases
$\sim 10^{41}$ erg in electromagnetic energy and high-energy particles (in the case of `spring' loops,
this is if the currents have opposing signs).  The high Lorentz factor achieved by this tiny explosion
allows a high efficiency of conversion to photons of frequency $\nu_{\rm peak} \sim c/2\pi {\cal R} 
\sim 50({\cal R}_{-1})^{-1}$ GHz, decreasing as $\sim \nu^{0.5-1}$ toward lower frequencies (Paper I).

We have identified two channels for LSD collisions, identifying non-repeating FRBs with rare
collisions in Galactic halos, whose rate is directly related to the the ratio ${\cal M}/{\cal R}$
(or equivalently to the string tension ${\cal T}$ in the `spring' model).  The collision
rate near SMBHs involves a more complicated chain of events, but we expect it to be much larger.

Equation (\ref{eq:nd}) shows that a SMBH could release O(1) electromagnetic burst each year even without
the accretion of additional LSD.   An even higher rate is possible if the SMBH experiences persistent accretion
at a low rate.
For example, if we restrict to black holes with masses less than or equal to the Galactic Center black
hole ($\sim 4\times 10^6\,M_\odot$), then Equation (\ref{eq:supply}) suggests that the rate per black 
hole is roughly $\sim 10^2$ per year for a time-averaged accretion rate $\dot m \sim 10^{-3}$.  
(This does not include an upward correction to the electromagnetic transient rate from gravitational 
lensing or reflection.)  Taking a cosmic density of such SMBHs of $10^{-2}$ Mpc$^{-3}$ \citep{cb10},
we deduce a collision rate $\sim 10^9$ Gpc$^{-3}$ yr$^{-1}$, some $\sim 10^4$-$10^5$ times the local non-repeating 
rate.

It should also be noted that a LSD cusp gathered around a SMBH would be disrupted by gravitational scattering
if the SMBH experienced a merger with another black hole of high or intermediate mass.  For this
reason, FRBs may avoid the cores of massive elliptical galaxies.  Although Equation (\ref{eq:nd}) predicts a
very high number of trapped LSDs around the most massive ($10^8$-$10^9\,M_\odot$) SMBHs, it is in such cases
that the probability of a major merger appears to be highest (e.g. \citealt{faber97}).


With the preceding caveat in mind, we conclude that nearby $\sim 10^{6-7}\,M_\odot $ SMBH are promising
targets for directed searches for 0.01-1 THz bursts.  These are predicted to be intrinsically brighter than
GHz FRBs, and would more easily avoid absorption.  We emphasize that most LSD collisions near SMBHs are
not expected to be detectable sources of GHz pulses:  even after accounting for the feedback of the strong wave
on ambient electrons, the optical depth to synchrotron absorption remains high in the GHz band.  This conclusion
holds (although not by a large margin) for the Galactic Center SMBH, with its low present accretion rate 
\citep{yuan12}. (See Equation (\ref{eq:nemax}) and Paper I.)  But the synchrotron optical depth is much 
smaller at 10-100 GHz.

We therefore highlight the prediction of a high rate of $\sim 100$ GHz bursts, mostly originating
near SMBHs, which are similar to FRBs but an order of magnitude brighter.  These pulses may be very 
narrow:   as limited by multi-path propagation some $\sim 10^{-4}$ times narrower than FRBs, corresponding
to sub-microsecond durations.

\subsection{Direct Detection of $\mu$Jy Radio Sources with \\
   Extreme Proper Motion}

The nearest LSD to the Earth is predicted to reside within our own Solar system: the mean separation between LSDs is 
$f_{\rm LSD}^{-1/3}(\rho_{\rm dm}/{\cal M})^{-1/3} \sim 35\,f_{\rm LSD}^{-1/3}{\cal M}_{20}^{-1/3}$ AU, implying a mean 
distance $\sim 20\,f_{\rm LSD}^{-1/3}{\cal M}_{20}^{-1/3}$ AU from the Sun.  A LSD magnetosphere
moving tangentially at a speed $\sim 300$ km s$^{-1}$ through a solar wind of proton density
$\sim 0.02$ cm$^{-3}$ and radial speed 350 km s$^{-1}$ would have a size $R_{\rm mag}
\sim 400{\cal M}_{20}^{1/6}{\cal R}_{-1}^{1/2}(a/20~{\rm AU})^{1/3}$ km from Equation (\ref{eq:rmag}) and angular 
diameter less than $0.1$ arcsec.  Its lifetime at 20 AU from the Sun would would be $\sim 0.3$ yr,
meaning that over a period of a few years the closest approach of a LSD to the Earth
and Sun would be $\sim 10$ AU, with magnetopause radius shrinking by a factor
$\sim 2^{-1/3}$ and angular size increasing by a factor $\sim 2^{2/3}$.   

Consider then the observability of coherent decametric or radio emission from the
LSD magnetosphere.  In the absence of a predictive theory of such emission, we simply
scale to planetary magnetospheres.  For example, a fraction $\sim 1\times 10^{-4}$ of the Solar wind
power incident on Jupiter's magnetosphere emerges as decametric emission, with
much higher efficiencies in the infrared band \citep{bhardwaj00}.  The Solar wind power incident
on an LSD magnetosphere would be
\ba
{\cal P}_w &\sim& {1\over 2}\rho_w V_w^3 \cdot \pi R_{\rm mag}^2  \nn
   &=& 3\times 10^{12}
\,{\cal M}_{20}^{1/3}{\cal R}_{-1}^{2/3} (a/20~{\rm AU})^{-4/3}\quad{\rm erg~s^{-1}}.\nn
\ea
The emission efficiency needed to produce a $\mu$Jy source in the GHz band at this distance
from the Earth is
\be
f_{\rm GHz} \sim 3.5\times 10^{-3}\,{\cal M}_{20}^{-1/3}{\cal R}_{-1}^{-2/3}\nu_9
(a/20~{\rm AU})^{10/3}.
\ee
Picking out a single sub-arcsecond source of this low luminosity and enormous proper motion
from the entire sky would certainly be challenging; but it would represent 
a direct detection of macroscopic dark matter and support for the ideas presented here.

\vfil\eject
\subsection{Constraints from Sub-Pixel Microlensing}\label{s:micro}

Microlensing of stars in M31 strongly constrains the
abundance of dark matter particles of masses exceeding $\sim 10^{20}$-$10^{21}$ g \citep{niikura17}
The lensing is detected below the pixel angular scale of a CCD detector \citep{crotts92}, and
the precise value of the mass threshold depends on details like the finite size of the lensed stars.  
Comparing with Equation (\ref{eq:Eem}) for the radiated energy in GHz photons, one sees that
this mass constraint is consistent with a `spring' tension ${\cal T} \sim 10^{-8}c^4/G$ (the value
that is needed to give a collision rate comparable to the observed FRB burst rate) and a conversion
efficiency $f_{\rm em} \sim 1$ of the mass of two colliding springs to the broadband electromagnetic
pulse.   

Collisional LSDs that are trapped near SMBH can combine to form a higher-energy tail that
can easily extend one-two orders of magnitude higher than the rest energy of the seed LSD
(Figure \ref{fig:energy_spectrum}).  If such a collisional population of LSDs were to dominate
the observed FRB rate, then the mass limit from microlensing could be easily avoided for
a larger tension $G{\cal T}/c^4 \sim 10^{-7}$.  In the approach to FRBs advanced here, this would be
the case for the repeating source FRB 121102.

Future microlensing searches will provide tight constraints on LSD progenitors of FRBs
given the small difference between the microlensing mass constraint and the minimum LSD mass
needed to power GHz FRB emission.

\begin{figure}
\epsscale{1.1}
\plotone{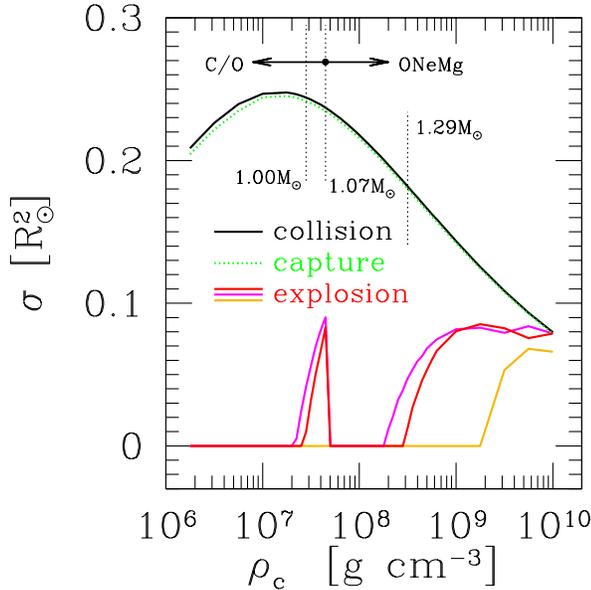}
\vskip .1in
\caption{Cross section for a LSD of tension ${\cal T} = 10^{-8}c^4/G$ and mass ${\cal M} = 4\times 10^{20}$ g,
moving at 200 km s$^{-1}$ at infinity, to interact with a WD of varying mass and central density.  
{\it Solid black curve:} collision.  {\it Dotted green curve:} collision and frictional trapping.  
{\it Red curve:} strong enough heating to trigger a deflagration/detonation, corresponding to 
$t_{\rm ign} < R_{\rm mag} / V_{\rm ps}$.  {\it Magenta curve:}  same as red curve, but for LSD of
mass ${\cal M} = 4\times 10^{19}$.  {\it Gold curve:} same as red curve, but for LSD of tension 
$G{\cal T}/c^4 = 10^{-9}$ and mass ${\cal M} = 4\times 10^{19}$ g (corresponding to an identical dipole size).}
\vskip .2in
\label{fig:cs_wd}
\end{figure}

\vfil\eject
\subsection{Induced Thermonuclear Explosions of \\ Massive ($> 1\,M_\odot$)
White Dwarf Stars}\label{s:wdc}

The interaction between LSDs and WD stars has observational consequences, involving
i) the triggering of hydrogen-depleted (Type Ia) supernova explosions;  ii) the enrichment of
galactic and intergalactic plasma with iron group elements; and iii) the depletion of the massive
tail of the WD population.  We now consider these effects in turn.

We found a threshold WD mass ($M_{\rm wd} = 1.0\,M_\odot$ for C/O 
and $1.29\,M_\odot$ for ONeMg composition) above which self-sustained thermonuclear burning would
be triggered by the infall of a LSD of tension ${\cal T} = 10^{-8}c^4/G$ (Figure \ref{fig:tcc}).  
The corresponding interaction
cross section is compared in Figure \ref{fig:cs_wd} with the total cross section for a direct
collision.   Reducing the LSD mass from $4\times 10^{20}$ g to $4\times 10^{19}$ g, and the
radius in proportion to the mass (corresponding to a fixed `spring' tension), slightly increases
the drag time (\ref{eq:tdrag}), and therefore the post-shock temperature during the deceleration
of the LSD.  As a result, explosions are slightly easier and the susceptible range of WD masses
widened slightly.  On the other hand, reducing the mass by the same amount but fixing the LSD radius
significantly reduces the drag time, and makes it harder to reach high temperatures.  In this
case, only the most massive ONeMg WDs develop self-sustained burning.

Recall also that burning is triggered relatively deeply in the C/O material (typically an enclosed mass below 
$\sim {1\over 2} M_{\rm wd}$), but in relatively shallow layers of the more massive WDs (Figure \ref{fig:trigger}).
This feature of the trigger mechanism is increasingly testable by measurements of the early light curves
of nearby Type Ia explosions, which probe the $^{56}$Ni yield in the outer ejecta shells
(e.g. SN 2011fe: \citealt{piro12}).

Combined nucleosynthetic and radiation transfer calculations of the detonations of $\sim 1\,M_\odot$
C/O WDs give encouraging light curves \citep{sim10} while avoiding the complications associated with
a deflagration-to-detonation transition. 
Nonetheless, as we discussed in Section \ref{s:prompt}, the shock induced by a LSD moving through a WD
appears to be a couple orders of magnitude too small to produce an unambiguous prompt detonation.  The
extent to which this conclusion depends on the extended (linear) geometry of the energy release needs
further examination.

\subsubsection{Lifetime of Local Massive White Dwarfs}\label{s:wd3}

Consider first the lifetime for a WD in the quoted range of masses.  Figure \ref{fig:cs_wd} shows the
cross section to induce a thermonuclear explosion, 
\be
\sigma_{\rm exp} = {\pi \ell_{\rm exp}^2\over (\Delta v)^2}.
\ee
Here $\ell_{\rm exp}$ is the corresponding angular momentum of the LSD particle, with respect to the target
WD.   In the case of $1.0-1.07\,M_\odot$ C/O WDs, we find that $\sigma_{\rm exp}$ averages to about $\sim 1/6$ 
the collision cross section $\sigma_{\rm col} = \pi \ell_{\rm col}^2/(\Delta v)^2$.

A target WD is assumed to reside in a rotating galactic disk that is immersed in a non-rotating and locally isothermal
dark matter halo with LSD mass fraction $f_{\rm LSD}$.  The distribution function of halo particles,
$f({\bf v}) \propto e^{-v^2/2\sigma^2}$, has a dispersion related to the circular speed by
\be
\sigma_{\rm LSD} = V_c/\sqrt{2}.
\ee
Averaging over the LSD velocity space ${\bf v}$, with $\Delta v = \sqrt{v_x^2 + v_y^2 + (v_z - V_c)^2}$, gives
\be
\left\langle\Delta v \sigma_{\rm exp}\right\rangle = \pi \ell_{\rm exp}^2\left\langle{1\over\Delta v}\right\rangle
   = {\pi\ell_{\rm exp}^2\over V_c}\,{\rm erf}\left({V_c\over \sqrt{2}\sigma}\right).
\ee
Here ${\rm erf}(1) = 0.843$.  

We obtain a mean lifetime for a nearby $1\,M_\odot$ WD to survive a collision with a LSD particle,
\ba\label{eq:tcol}
t_{\rm col}(1\,M_\odot) &=& {{\cal M}\over f_{\rm LSD}\rho_{\rm dm} \langle \Delta v \sigma_{\rm exp}\rangle}\nn
&=& 1.8\times 10^9\,f_{\rm LSD}^{-1}{\cal M}_{20}\left({\sigma_{\rm col}/6\over \sigma_{\rm exp}}\right)\quad {\rm yr}.\nn
\ea
The lifetime of massive ONeMg WDs is roughly half this (Figure \ref{fig:cs_wd}).
Here we have used values for the local Milky Way dark matter density, $\rho_{\rm dm} \simeq 7\times 10^{-25}$ g cm$^{-3}$,
and circular speed $V_c = 230$ km s$^{-1}$, taken from \cite{mcmillan17}.  

If LSDs comprise a large fraction of the Galactic dark matter, collisions with them
produce a strong depletion of nearby cold and massive WDs in the mass intervals $1.0-1.07\,M_\odot$ and
$> 1.29\,M_\odot$.  This mass function is not yet accurately measured,
but it provides strong potential tests of the mechanism of WD disruption described here.
Evidence for a depletion of massive WDs closer than 20 pc to the Sun has been discussed by \cite{tremblay16}, whereas
evidence for a feature near $1\,M_\odot$ in the mass spectrum of local WDs cooler than 12,000 $^\circ$K has been
described by \cite{rebassa15}.   

The lifetime of a massive WD will depend on its position within its host galaxy, decreasing toward the center.
Massive white dwarfs also appear to survive in binary systems, but estimates of their ages are made
ambiguous by the possibility of mass growth by accretion, and so we do not consider the constraints imposed
by their existence here.

\subsubsection{Global Rate of LSD-Induced WD Explosions}

If the LSD mass is in the range (${\cal M} \sim 10^{20}$ g) we find is necessary to power FRBs of 
energy $10^{39}$ erg, WDs in quoted mass range will have lifetimes significantly shorter than the age
of the universe:  a large fraction are eventually destroyed by collision with a LSD.  

This leads to a well defined prediction for the number of Type Ia supernovae per unit stellar mass, averaged
over cosmic volume.  We take a \cite{kroupa01} stellar initial  mass function (IMF) $dN/dM_\star$, 
\ba\label{eq:kroupa}
{dN\over dM_\star} = \left\{
     \begin{array}{ll}                                                    
      (K_1/M_{\star 1}) (M_\star/M_{\star 1})^{-\alpha_1}  \quad & (M_\star < M_{\star 1}) \\
      (K_1/M_{\star 1}) (M_\star/M_{\star 1})^{-\alpha_2}  \quad & (M_{\star 1} < M_\star < M_{\star 2}) \\
      (K_3/M_{\star 2}) (M_\star/M_{\star 2})^{-\alpha_3}  \quad & (M_\star > M_{\star 2})
     \end{array}
   \right.\nn
\ea
Here $M_{\star 1} = 0.08\,M_\odot$, $M_{\star 2} = 0.5\,M_\odot$, $\alpha_1 = 0.3$, $\alpha_2 = 1.7$ and $\alpha_3
= 2.3$.
The coefficients $K_i$ are related to the total stellar mass by $K_1 = 0.77\,M_{\star,\rm tot}/M_\odot = 1.73 K_3$.

To obtain an explosion rate, we make use of the relation between zero-age main sequence stellar mass and 
WD mass obtained from state-of-the-art stellar evolution calculations \citep{doherty15}.   All numbers quoted
here refer to solar metallicity.  A star of mass $M_\star > M_{\star,1} = 5.8\,M_\odot$ is needed to 
produce a $1\,M_\odot$ WD, increasing to $M_{\star,2} = 6.4\,M_\odot$ for the highest mass C/O WDs ($1.07\,M_\odot$).
We neglect the possibility of hybrid carbon-ONeMg WDs here.  Finally ONeMg WDs with mass exceeding
$1.29\,M_\odot$ form from $7.9-8.5\,M_\odot$ progenitors.  
Integrating over the initial progenitor mass for C/O WDs heavier than $1\,M_\odot$, we find
\be\label{eq:nIa}
N_{\rm Ia}({\rm C/O}) = \int_{M_{\star,1}}^{M_{\star,2}} dM_\star {dN\over dM_\star} = 1.7\times 10^{-3} 
{M_{\star,\rm tot}\over M_\odot}.
\ee
The corresponding yield of ONeMg WDs more massive than $1.29\,M_\odot$ is
\be\label{eq:nIab}
N_{\rm Ia}({\rm ONeMg}) = 8.5\times 10^{-4}{M_{\star,\rm tot}\over M_\odot}.
\ee

Deductions of the Type Ia SN rate from measurements of iron enrichment in the hot plasma of rich galaxy clusters
suggests a normalization $N_{\rm Ia} \simeq 3.4\times 10^{-3}\,M_{\star,\rm tot}/M_\odot$, based on a $^{56}{\rm Fe}$ yield
of $0.7\,M_\odot$ per explosion;  whereas the observed rates of Type Ia supernova in galaxy clusters suggest values 
$\sim 2-2.5\times 10^{-3}\,M_{\star,\rm tot}/M_\odot$ \citep{mm12}.

The approximate agreement of Equations (\ref{eq:nIa}) and (\ref{eq:nIab})
with these empirical values is encouraging: our evaluation of the
Type Ia rate depends only a i) a prescribed WD mass range within which LSD collisions produce explosions
and ii) the stellar IMF.    Furthermore,
this triggering mechanism is agnostic as to the situation of the target WD, which could be a single star
as well one accompanied by a degenerate or non-degenerate companion.  Type Ia supernovae are predicted to
arise from all types of WD binaries with a frequency proportional to their formation rate, at least to the
extent that these rates do not depend on the local stellar or dark matter density.


\subsection{Constraints on `Spring' Tension from CMB Lensing and Low-Frequency Gravity Wave Emission}\label{s:gw}

The concrete microphysical example of LSDs given here involves macroscopic loops of GUT-scale
superconducting strings.  The spread in loop masses is restricted to lie within a decade centered
around the value (${\cal M} \sim 10^{20}$ g) needed to power FRBs.  If loops a decade or two smaller 
made a significant contribution to the cold dark matter density, then WDs with masses
exceeding $\sim 1\,M_\odot$  would have a lifetime shorter than $\sim 10^7$-$10^8$ yrs.  A significant
density in loops with masses exceeding $\sim 10^{21}$ g appears to be inconsistent with the absence
of short-timescale microlensing of stars in M31 \citep{niikura17}.

At first sight, such a peaked distribution of loop sizes is not consistent with the approach to cosmic
string formation initiated by \cite{kibble76}.  This involves an infinite network of string
combined with closed loops that break off the string during its self-intersection.  This network
has no characteristic scale smaller than the comoving horizon size during the symmetry breaking phase
transition that created the string network.   In Section \ref{s:preheat} we make a connection between
$\sim$ mm-sized LSDs and low-temperature post-inflationary preheating of the universe.

The long string which might be present plays no role
in the FRB emission process described here and in Paper I, but would have observational consequences
involving i) gravity wave emission \citep{vilenkin81,turok84,vv85}; ii) gravitational lensing of background galaxies
\citep{vilenkin84,ks84,gott85}; and gravitational lensing of the cosmic microwave background (CMB)
\citep{planck14,lazanu15,lizarraga16}.  Kink-like structure on the string and loops with redshifted periods 
of $P_{\rm gw} \sim 1$ year have a size limited by gravity-wave emission.  The emission redshift is determined
from 
\be
{G{\cal T}\over c^4} ct(z_{\rm em}) \sim {cP_{\rm gw}\over 1+z_{\rm em}}
\ee
and is relatively small, $1+z_{\rm em} \sim 10^4\,(G{\cal T}/c^4)_{-8}$.   

As a result, any long-string network associated with the LSD loops should lens the CMB and
produce gravitational radiation in the standard amount.   An upper bound $G{\cal T}/c^4 < 1.3\times 10^{-7}$
has been derived by combining data from the Planck satellite and from the high-$\ell$ microwave background 
experiments ACT and SPT \citep{planck14}.  A nearly identical constraint is derived by combining these data
with string network simulations \citep{lazanu15,lizarraga16}.  The most conservative current bounds from pulsar timing are 
$G{\cal T}/c^4 \lesssim 1.3\times 10^{-7}$ from the EPTA \citep{lentati15} and  $G{\cal T}/c^4 < 3.3\times 10^{-8}$
from the NANOGrav experiment \citep{arzoumanian16}.  These constraints allow the size of string loops to be freely 
variable; a much tighter constraint, $G{\cal T}/c^4 < 1.3\times 10^{-10}$ \citep{arzoumanian16}, is 
derived from a more conventional estimate of loop size as determined by self-intersection and decay by the
emission of gravitational waves.  The CMB constraints and most conservative versions of the pulsar timing
constraints are therefore consistent with a `spring' tension ${\cal T} \sim 10^{-8}c^4/G$, but it is apparent
that significant departures from the standard string network calculations must be present to accomodate
more fully the pulsar timing results.

Finally, we note that strong gravitational lensing by the long string would occur on a scale 
$8\pi G{\cal T}/c^4 = 0.05 (G{\cal T}/c^4)_{-8}$ arcsec.  This angular scale is small enough to have
been missed so far by observational searches for strong lenses.

\vskip .2in
\subsection{Formation During Post-Inflationary Preheating}\label{s:preheat}

When considering the origin of macroscopic field structures in the early universe, two useful reference
points are i) the mass-energy contained with a horizon-sized volume at the epoch of their formation;
and ii) the mass of dark matter extrapolated backward in
the cosmic expansion to this volume.  Our focus here is on the electroweak scale.  The temperature
of electroweak symmetry restoration implied by a minimal standard model Higgs particle of mass 
125 GeV is $\kB T_{\rm EW} \simeq 165$ GeV \citep{dine92}, 
at a redshift $z_{\rm EW} = (T_{\rm EW}/T_{\rm CMB}) \times [g_{*,\rm EW}/3]^{1/3} = 2.3\times
10^{15}$, where $g_{*,\rm EW} \simeq 107$ is the number of relativistic degrees of freedom at 
this temperature.
One finds 
\be
M_{\rm hor}(z) = \left({c\over H}\right)^3 {\pi^2 g_*\over 30\hbar^3 c^5} (\kB T)^4  
\sim 10^{27}\left({z\over z_{\rm EW}}\right)^{-2}\; {\rm g},
\ee
where $H$ is the Hubble parameter at redshift $z < z_{\rm EW}$.
By comparison, the cold dark matter mass is
\be\label{eq:mdhor}
M_{\rm d,H}(z) = \left({c\over H}\right)^3 (1+z)^3
 \rho_d(0) \sim 5\times 10^{15}\left({z\over z_{\rm EW}}\right)^{-3}\; {\rm g}.
\ee

The LSD mass ${\cal M} \sim 10^{20}$ g considered here is intermediate between these two values.
That means that Friedmann-type expansion through the electroweak phase transition
can only produce stable macroscopic field structures of a much smaller mass.  On the other hand,
higher LSD masses are possible if the entropy/dark matter ratio increases with time, which
would be the case if inflationary preheating occurred at a temperature below $T_{\rm EW}$.  Equating
expression (\ref{eq:mdhor}) with ${\cal M}$ gives a preheating temperature $T_{\rm pr} \sim 0.03\,T_{\rm EW}$.

\begin{appendix}

\section{Collisional Cross Section of Free Classical Dipoles}

The equations describing the collision between two dipoles ($i,j = 1,2$ and $i \neq j$) are
\ba\label{eq:dipoleq}
{\cal M}_r {d{\bf v}_{ij}\over dt} &=& {\partial[\bmu_i\cdot{\bf B}_j({\bf r}_{ij})]\over\partial{\bf r}_i}
  = -{\partial[\bmu_j\cdot{\bf B}_i({\bf r}_{ij})]\over\partial{\bf r}_j}; \nn
{d\bmu_i\over dt} &=& \bOmega_i\times\bmu_i; \nn
{\cal I}_i {d\bOmega_i\over dt} &=& \bmu_i\times{\bf B}_j({\bf r}_{ij}); \quad \bOmega_i\cdot {\bf B}_j({\bf r}_{ij}) = {\rm const};\nn
{\bf r}_{ij} &=& r_{ij}\hat r_{ij} = {\bf r}_i - {\bf r}_j; \quad {\bf v}_{ij} = {d{\bf r}_{ij}\over dt}.
\ea
Here ${\cal M}_r = {\cal M}_1{\cal M}_2/({\cal M}_1 + {\cal M}_2)$ is
the reduced mass, ${\bf B}_i$ is the magnetic field sourced by dipole $i$, and
\be
-{\bmu}_1\cdot {\bf B}_2({\bf r}_{12}) = -{\bmu}_2\cdot {\bf B}_1({\bf r}_{21}) =
 {\bmu_1\cdot\bmu_2 - 3(\bmu_1\cdot\hat r_{12})(\bmu_2\cdot\hat r_{12}) \over r_{12}^3}
\ee
is the interaction energy.  Scaling distances to $R_{\rm col} = (2\mu_1\mu_2/{\cal M}_r (\Delta v)^2)^{1/3}$,
and the time to $R_{\rm col}/\Delta v$, one finds
\be\label{eq:eom}
{R_{\rm col}\over \Delta v^2}{d{\bf v}_{ij}\over dt} = {3\over 2}\left({R_{\rm col}\over r_{12}}\right)^4
   \biggl[ \hat\mu_i(\hat\mu_j\cdot\hat r_{ij}) + \hat\mu_j(\hat\mu_i\cdot\hat r_{ij}) 
           + \hat r_{ij}(\hat\mu_i\cdot\hat\mu_j -  5\hat\mu_i\cdot\hat{r}_{ij}\hat\mu_j\cdot\hat{r}_{ij}) \biggr].
\ee
We also scale the two rotational moments of inertia to a characteristic
value, ${\cal I}_i = \widetilde{\cal I}_i {\cal I}$, and the angular
velocity to $\bOmega_i = (\Delta v/R_{\rm col})\,f\,\hat\Omega_i$,
where $f  = ({\cal M}_r R_{\rm col}^2/2{\cal I})^{1/2}$ is the
fastness parameter.  This gives
\ba\label{eq:eom2}
{R_{\rm col}\over\Delta v}{d\hat\mu_i\over dt} &=& 
f\left(\hat\Omega_i\times \hat\mu_i\right);\nn
\widetilde{\cal I}_i{R_{\rm col}\over\Delta v}{d\hat\Omega_i\over d t} &=&
       f\left({R_{\rm col}\over r_{12}}\right)^3\,
\hat\mu_i\times \Bigl[3(\hat\mu_j\cdot\hat r_{ij})\hat r_{ij} - \hat\mu_j\Bigr].
\ea

The Equations (\ref{eq:eom}), (\ref{eq:eom2}) are solved for a large number of randomly chosen initial orientations of the
two colliding dipoles to obtain the cross section shown in Figure \ref{fig:cs}.

\end{appendix}


\begin{thebibliography}{}
\bibitem[Arzoumanian et al.(2016)]{arzoumanian16} Arzoumanian, Z., Brazier, A., Burke-Spolaor, S., et al.\ 2016, \apj, 821, 13 
\bibitem[Bailey \& Hiatt(1972)]{bh72} Bailey, A.~B., \& Hiatt, J.\ 1972, AIAA Journal, 10, 1436 
\bibitem[Begelman et al.(2006)]{begelman06} Begelman, M.~C., Volonteri, M., \& Rees, M.~J.\ 2006, \mnras, 370, 289 
\bibitem[Bhardwaj \& Gladstone(2000)]{bhardwaj00} Bhardwaj, A., \& Gladstone, G.~R.\ 2000, Reviews of Geophysics, 38, 295 
\bibitem[Binney \& Tremaine(2008)]{bt08} Binney, J., \& Tremaine, S.\ 2008, Galactic Dynamics: Second Edition,
Princeton University Press, Princeton
\bibitem[Blandford(1977)]{blandford77} Blandford, R.~D.\ 1977, \mnras, 181, 489 
\bibitem[Blandford \& Begelman(1999)]{bb99} Blandford, R.~D., \& Begelman, M.~C.\ 1999, \mnras, 303, L1 
\bibitem[Capela et al.(2013)]{capela13} Capela, F., Pshirkov, M., \& Tinyakov, P.\ 2013, \prd, 87, 023507 
\bibitem[Capela et al.(2014)]{capela14} Capela, F., Pshirkov, M., \& Tinyakov, P.\ 2014, \prd, 90, 083507 
\bibitem[Caramete \& Biermann(2010)]{cb10} Caramete, L.~I., \& Biermann, P.~L.\ 2010, \aap, 521, A55 
\bibitem[Caughlan \& Fowler(1988)]{caughlan88} Caughlan, G.~R., \& Fowler, W.~A.\ 1988, Atomic Data and Nuclear Data Tables, 40, 283 
\bibitem[Chatterjee et al.(2017)]{chatterjee17} Chatterjee, S., Law, C.~J., Wharton, R.~S., et al.\ 2017, \nat, 541, 58 
\bibitem[Choi et al.(2013)]{choi13} Choi, J.-H., Shlosman, I., \& Begelman, M.~C.\ 2013, \apj, 774, 149 
\bibitem[Choi et al.(2015)]{choi15} Choi, J.-H., Shlosman, I., \& Begelman, M.~C.\ 2015, \mnras, 450, 4411 
\bibitem[Churazov et al.(2017)]{churazov17} Churazov, E., Khabibullin, I., Sunyaev, R., \& Ponti, G.\ 2017, \mnras, 465, 45 
\bibitem[Copeland et al.(1987)]{copeland87} Copeland, E., Turok, N., \& Hindmarsh, M.\ 1987, Physical Review Letters, 58, 1910 
\bibitem[Crawford et al.(2016)]{crawford16} Crawford, F., Rane, A., Tran, L., et al.\ 2016, \mnras, 460, 3370 
\bibitem[Crotts(1992)]{crotts92} Crotts, A.~P.~S.\ 1992, \apjl, 399, L43 
\bibitem[Davis \& Shellard(1989)]{davis89} Davis, R.~L., \& Shellard, E.~P.~S.\ 1989, Nuclear Physics B, 323, 209 
\bibitem[Dine et al.(1992)]{dine92} Dine, M., Leigh, R.~G., Huet, P., Linde, A., \& Linde, D.\ 1992, \prd, 46, 550 
\bibitem[Doherty et al.(2015)]{doherty15} Doherty, C.~L., Gil-Pons, P., Siess, L., Lattanzio, J.~C., and 
                              Lau, H.~H.~B.\ 2015, \mnras, 446, 2599
\bibitem[Dursi \& Timmes(2006)]{dursi06} Dursi, L.~J., \& Timmes, F.~X.\ 2006, \apj, 641, 1071 
\bibitem[Faber et al.(1997)]{faber97} Faber, S.~M., Tremaine, S., Ajhar, E.~A., et al.\ 1997, \aj, 114, 1771 
\bibitem[Farmer et al.(2015)]{farmer15} Farmer, R., Fields, C.~E., \& Timmes, F.~X.\ 2015, \apj, 807, 184 
\bibitem[Fattahi et al.(2016)]{fattahi16} Fattahi, A., Navarro, J.~F., Sawala, T., et al.\ 2016, arXiv:1607.06479 
\bibitem[Gondolo \& Silk(1999)]{gondolo99} Gondolo, P., \& Silk, J.\ 1999, Physical Review Letters, 83, 1719 
\bibitem[Gott(1985)]{gott85} Gott, J.~R., III 1985, \apj, 288, 422 
\bibitem[Governato et al.(2012)]{governato12} Governato, F., Zolotov, A., Pontzen, A., et al.\ 2012, \mnras, 422, 1231 
\bibitem[Graham et al.(2015)]{graham15} Graham, P.~W., Rajendran, S., \& Varela, J.\ 2015, \prd, 92, 063007 
\bibitem[G{\"u}ltekin et al.(2009)]{gultekin09} G{\"u}ltekin, K., Richstone, D.~O., Gebhardt, K., et al.\ 2009, \apj, 
698, 198 
\bibitem[Haws et al.(1988)]{haws88} Haws, D., Hindmarsh, M., \& Turok, N.\ 1988, Physics Letters B, 209, 255 
\bibitem[Kaiser \& Stebbins(1984)]{ks84} Kaiser, N., \& Stebbins, A.\ 1984, \nat, 310, 391 
\bibitem[Katz(2016)]{katz16} Katz, J.~I.\ 2016, arXiv:1611.01243 
\bibitem[Kibble(1976)]{kibble76} Kibble, T.~W.~B.\ 1976, Journal of Physics A Mathematical General, 9, 1387 
\bibitem[Klypin et al.(2011)]{klypin11} Klypin, A.~A., Trujillo-Gomez, S., \& Primack, J.\ 2011, \apj, 740, 102 
\bibitem[Kroupa(2001)]{kroupa01} Kroupa, P.\ 2001, \mnras, 322, 231 
\bibitem[Lazanu et al.(2015)]{lazanu15} Lazanu, A., Shellard, E.~P.~S., \& Landriau, M.\ 2015, \prd, 91, 083519 
\bibitem[Lentati et al.(2015)]{lentati15} Lentati, L., Taylor, S.~R., Mingarelli, C.~M.~F., et al.\ 2015, \mnras, 453, 2576 
\bibitem[Lizarraga et al.(2016)]{lizarraga16} Lizarraga, J., Urrestilla, J., Daverio, D., Hindmarsh, M., \& Kunz, M.\ 2016, Journal of Cosmology and Astroparticle Phys., 10, 042 
\bibitem[Loeb \& Rasio(1994)]{loeb94} Loeb, A., \& Rasio, F.~A.\ 1994, \apj, 432, 52 
\bibitem[Lorimer et al.(2007)]{lorimer07} Lorimer, D.~R., Bailes, M., McLaughlin, M.~A., Narkevic, D.~J., \& Crawford, F.\ 2007, Science, 318, 777 
\bibitem[Luan \& Goldreich(2014)]{lg14} Luan, J., \& Goldreich, P.\ 2014, \apjl, 785, L26 
\bibitem[Maoz \& Mannucci(2012)]{mm12} Maoz, D., \& Mannucci, F.\ 2012, P.A.S.A., 29, 447 
\bibitem[Marcote et al.(2017)]{marcote17} Marcote, B., Paragi, Z., Hessels, J.~W.~T., et al.\ 2017, \apjl, 834, L8 
\bibitem[McMillan(2017)]{mcmillan17} McMillan, P.~J.\ 2017, \mnras, 465, 76 
\bibitem[Nayakshin(2003)]{nayakshin03} Nayakshin, S.\ 2003, Astronomische Nachrichten Supplement, 324, 483 
\bibitem[Navarro et al.(1997)]{nfw97} Navarro, J.~F., Frenk, C.~S., \& White, S.~D.~M.\ 1997, \apj, 490, 493 
\bibitem[Niemeyer \& Woosley(1997)]{niemeyer97} Niemeyer, J.~C., \& Woosley, S.~E.\ 1997, \apj, 475, 740 
\bibitem[Niikura et al.(2017)]{niikura17} Niikura, H., Takada, M., Yasuda, N., et al.\ 2017, arXiv:1701.02151 
\bibitem[Novikov \& Thorne(1973)]{nt73} Novikov, I.~D., \& Thorne, K.~S.\ 1973, in Black Holes, eds. C. DeWitt and B. DeWitt (Paris:  Gordon and Breach), pp. 343-450
\bibitem[Ostriker et al.(1986)]{otw86} Ostriker, J.~P., Thompson, C., \& Witten, E.\ 1986, Physics Letters B, 180, 231 
\bibitem[Peebles(1972)]{peebles72} Peebles, P.~J.~E.\ 1972, General Relativity and Gravitation, 3, 63 
\bibitem[Piro(2012)]{piro12} Piro, A.~L.\ 2012, \apj, 759, 83 
\bibitem[Planck Collaboration et al.(2014)]{planck14} Planck Collaboration, Ade, P.~A.~R., Aghanim, N., etal.\ 2014, \aap, 571, A25 
\bibitem[Planck Collaboration et al.(2016)]{planck16} Planck Collaboration, Ade, P.~A.~R., Aghanim, N., etal.\ 2016, \aap, 594, A13 
\bibitem[Prada et al.(2012)]{prada12} Prada, F., Klypin, A.~A., Cuesta, A.~J., Betancort-Rijo, J.~E., \& Primack, J.\ 2012,
\mnras, 423, 3018 
\bibitem[Ravi et al.(2016)]{ravi16} Ravi, V., Shannon, R.~M., Bailes, M., et al.\ 2016, Science, 354, 1249 
\bibitem[Rebassa-Mansergas et al.(2015)]{rebassa15} Rebassa-Mansergas, A., Rybicka, M., Liu, X.-W., Han, Z., \& Garc{\'{\i}}a-Berro, E.\ 2015, \mnras, 452, 1637 
\bibitem[Schaller et al.(2016)]{schaller16} Schaller, M., Frenk, C.~S., Theuns, T., et al.\ 2016, \mnras, 455, 4442 
\bibitem[Scholz et al.(2016)]{scholz16} Scholz, P., Spitler, L.~G., Hessels, J.~W.~T., et al.\ 2016, \apj, 833, 177 
\bibitem[Shakura \& Sunyaev(1973)]{ss73} Shakura, N.~I., \& Sunyaev, R.~A.\ 1973, \aap, 24, 337 
\bibitem[Sim et al.(2010)]{sim10} Sim, S.~A., R{\"o}pke, F.~K., Hillebrandt, W., et al.\ 2010, \apjl, 714, L52 
\bibitem[Spitler et al.(2016)]{spitler16} Spitler, L.~G., Scholz, P., Hessels, J.~W.~T., et al.\ 2016, \nat, 531, 202 
\bibitem[Steigman et al.(1978)]{steigman78} Steigman, G., Sarazin, C.~L., Quintana, H., \& Faulkner, J.\ 1978, \aj, 83, 1050 
\bibitem[Tendulkar et al.(2017)]{tendulkar17} Tendulkar, S.~P., Bassa, C.~G., Cordes, J.~M., et al.\ 2017, \apjl, 834, L7 
\bibitem[Thornton et al.(2013)]{thornton13} Thornton, D., Stappers, B., Bailes, M., et al.\ 2013, Science, 341, 53 
\bibitem[Thompson(2017)]{t17} Thompson, C. \apj, submitted, arXiv:1703.00393
\bibitem[Timmes \& Swesty(2000)]{ts00} Timmes, F.~X., \& Swesty, F.~D.\ 2000, \apjs, 126, 501 
\bibitem[Tremblay et al.(2016)]{tremblay16} Tremblay, P.-E., Cummings, J., Kalirai, J.~S., et al.\ 2016, \mnras, 461, 2100 
\bibitem[Turok(1984)]{turok84} Turok, N.\ 1984, Nuclear Physics B, 242, 520 
\bibitem[Vachaspati \& Vilenkin(1985)]{vv85} Vachaspati, T., \& Vilenkin, A.\ 1985, \prd, 31, 3052 
\bibitem[Vilenkin(1981)]{vilenkin81} Vilenkin, A.\ 1981, Physics Letters B, 107, 47 
\bibitem[Vilenkin(1984)]{vilenkin84} Vilenkin, A.\ 1984, \apjl, 282, L51 
\bibitem[Wang \& Merritt(2004)]{wang04} Wang, J., \& Merritt, D.\ 2004, \apj, 600, 149 
\bibitem[Witten(1985)]{witten85} Witten, E.\ 1985, Nuclear Physics B, 249, 557 
\bibitem[Young(1980)]{young80} Young, P.\ 1980, \apj, 242, 1232 
\bibitem[Yu et al.(2014)]{yu14} Yu, Y.-W., Cheng, K.-S., Shiu, G., \& Tye, H.\ 2014, J. Cosmol. Astropart. Phys., 11, 040 
\bibitem[Yuan et al.(2003)]{yuan03} Yuan, F., Quataert, E., \& Narayan, R.\ 2003, \apj, 598, 301 
\bibitem[Yuan et al.(2012)]{yuan12} Yuan, F., Wu, M., \& Bu, D.\ 2012, \apj, 761, 129 


\end{thebibliography}
\end{document}